%% file: root.tex
\documentclass[conference]{IEEEtran}
\IEEEoverridecommandlockouts
\usepackage{cite}
\usepackage{amsmath,amssymb,amsfonts,newtxmath,mathtools}
\usepackage{algorithm}
\usepackage{graphicx}
\usepackage{textcomp}
\usepackage{algpseudocode}
\algtext*{EndFor} 
\algtext*{EndIf}  
\algdef{SE}[FOR]{For}{EndFor}[1]
  {\algorithmicfor\ #1}
  {\algorithmicend\ \algorithmicfor}%

\algdef{SE}[IF]{If}{EndIf}[1]
  {\algorithmicif\ #1\ \algorithmicthen}
  {\algorithmicend\ \algorithmicif}%
\usepackage{xcolor}
\usepackage{float}
\usepackage{framed}
\usepackage{wrapfig}
\usepackage{enumerate}
\usepackage{caption}
\usepackage{rotating}
\usepackage{subcaption}
\usepackage{adjustbox}
\usepackage{multirow}
\usepackage{array}
\usepackage{tabularx}
\usepackage{booktabs}
\usepackage{nameref}
\usepackage{ifthen}
\usepackage{makecell}
\usepackage{cancel}
\usepackage{soul}
\usepackage{ulem}
\usepackage{titlesec}
\newboolean{longversion}
\usepackage{titlesec}
\setboolean{longversion}{true} 
\newcommand{\versionedText}[2]{\ifthenelse{\boolean{longversion}}{#1}{#1 #2}}
\DeclareCaptionLabelFormat{figure}{Figure #2}
\def\BibTeX{{\rm B\kern-.05em{\sc i\kern-.025em b}\kern-.08em
    T\kern-.1667em\lower.7ex\hbox{E}\kern-.125emX}}
    
\renewcommand{\thetable}{\arabic{table}} 
\definecolor{barakcolor}{RGB}{255, 0, 0}

\begin{document}

\title{Data-Driven Cellular Network Selector for Vehicle Teleoperations\thanks{This research was partially supported by the Council For Higher Education - Planning and Budgeting Committee.}
}

\author{
\IEEEauthorblockN{
Barak Gahtan\IEEEauthorrefmark{1},  
Reuven Cohen\IEEEauthorrefmark{1}, 
Alex M. Bronstein\IEEEauthorrefmark{1},
Eli Shapira\IEEEauthorrefmark{2}
}
\IEEEauthorblockA{\IEEEauthorrefmark{1}Technion Israel Institute of Technology, Haifa, Israel\\
Email: \{barakgahtan, rcohen, bron\}@cs.technion.ac.il}
\IEEEauthorblockA{\IEEEauthorrefmark{2}DriveU Ltd., Email: eli@driveu.auto}
}
\maketitle
\begin{abstract}
Remote control of robotic systems, also known as teleoperation, is crucial for the development of autonomous vehicle (AV) technology. It allows a remote operator to view live video from AVs and, in some cases, to make real-time decisions. The effectiveness of video-based teleoperation systems is heavily influenced by the quality of the cellular network and, in particular, its packet loss rate and latency. To optimize these parameters, an AV can be connected to multiple cellular networks and determine in real time over which cellular network each video packet will be transmitted. We present an algorithm, called Active Network Selector (ANS), which uses a time series machine learning approach for solving this problem. We compare ANS to a baseline non-learning algorithm, which is used today in commercial systems, and show that ANS performs much better, with respect to both packet loss and packet latency.
\end{abstract}

\section{Introduction}\label{INTRO}
\input{sections/1-introduction}
\section{Related Work}\label{RELATEDSEC}
\input{sections/2-related-work}


\section{A Deep Learning Forecasting Framework}\label{methodology}
\input{sections/4-methodology}

\section{Machine Learning Models Training Results}\label{training-res}
\input{sections/5-training-results}

\section{Evaluation on New Test Drives}\label{eval-res}
\input{sections/6-evaluation-results}

\section{An Algorithm for Network Selection}\label{ppNS}
\input{sections/8-ppNS}

\section{Conclusions}\label{conclusion}
\input{sections/7-conclusion}

\begin{tiny}
\bibliographystyle{plain}
\bibliography{References}
\end{tiny}
\end{document}

%% file: sections/1-introduction.tex
Remote control of robotic systems and machines, also known as teleoperation, plays a key role in AV technology. The idea behind AV teleoperation is that a remote operator in a control center can intervene in the operation of one or more AVs, ranging from direct driving to acting as a remote safety driver in challenging situations \cite{Teleoperation}. The overarching objective is, however, to provide AVs with sufficient high-quality data and training so that they can make intelligent decisions themselves in real-time, eliminating the need for a remote driver. But even if the AV becomes fully autonomous, it will have to send high-quality real-time video to a control center. 

Acquiring high-quality live video over cellular networks with latency not exceeding 100ms is one of the greatest challenges of teleoperation. To address the formidable latency constraints, UDP-based (rather than TCP-based) transport is used. To address possible packet losses, forward error correction (FEC) techniques, which use cross-packet redundancy, are employed \cite{GOLAGHAZADEH2022116597, tsai2010sub}.

Cellular networks are known to be unstable and exhibit dynamic behavior for many reasons. Radio interference, congestion, and handover between base stations all undermine network stability and make it very challenging to operate on. Even with the most advanced FEC-based video encoding, the quality of the live video from the AV is heavily influenced by the quality of the cellular network and, in particular, its packet loss rate and latency. To minimize these parameters, it is possible to connect AVs to multiple cellular networks. The AV is then required to determine in real-time over which cellular network its next video packets should be transmitted such that the packet loss rate and latency are minimized. We call this the multi-cellular packet routing (MPR) problem. To the best of our knowledge, we are the first to address this problem.

Transmitting each packet over multiple cellular networks increases the complexity of the system, the communication cost, and requires a new family of transport protocols. This paper seeks a less complex method -- transmitting video packets over one cellular network -- for solving the MPR problem.
\begin{figure}[t]
    \centering
    \includegraphics[width=1\columnwidth]{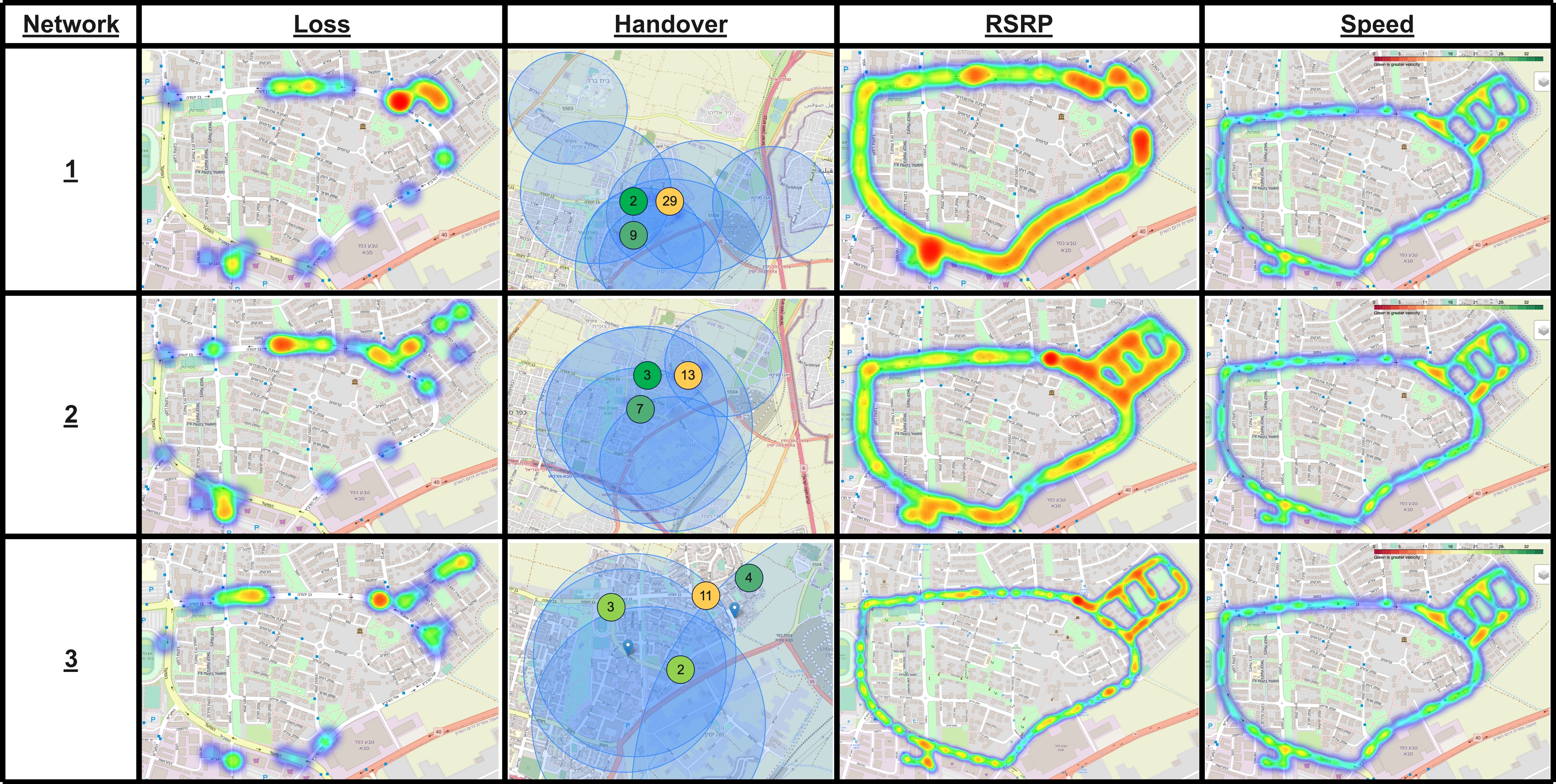}
    \caption{The results of our communication test drives: Red spots indicate high packet loss in the Loss column and poor signal quality in the RSRP column; green spots indicate high velocity in the Speed column; the Handover column indicates the number of handover events that occurred in each location}
    \label{fig:drive}
\end{figure}
We conducted test drives with a vehicle connected to three different cellular networks. Figure \ref{fig:drive} shows the correlation between packet losses, handovers, user measurement of the received signal quality from the cell (RSRP), and vehicle speed. The first column on the left (Loss) is a heat map that shows the number of lost packets on each cellular network and in each location of the moving vehicle. The red color indicates a higher packet loss. The second column (Handover) is a heat map that shows the locations of handover events. The numbers on these maps indicate the total number of handover events that occur in each location. The blue circles in this column indicate the antenna coverage as documented in the OpenCelliD database \cite{OpenCellID}. The third column (RSRP) shows the user measurements of the received signal quality from the cell, represented by the reference signal received power (red indicates low quality signal). Finally, the last column is a heat map of vehicle velocity at various locations, (green indicates high velocity). The most important observation from this figure is that there is a strong correlation between the RSRP quality, the frequency of handover events, and the packet loss rate. For example, for network 1 in this figure, when the RSRP measurements are closer to dark red, the number of packet losses and the number of handovers are greater.

\begin{figure}[t]
    \centering
    \includegraphics[width=1\columnwidth]{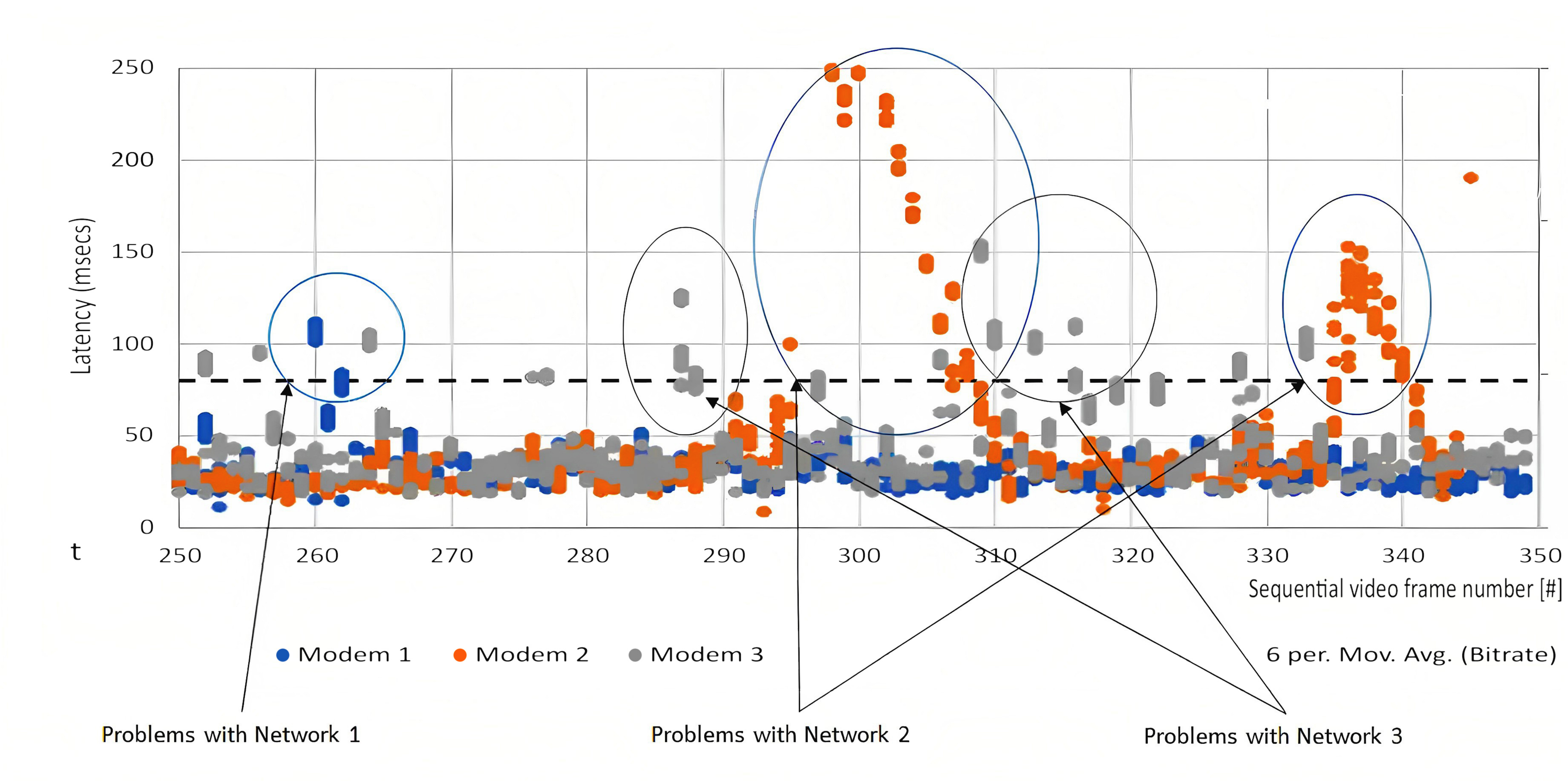}
    \caption{Latency comparison of three different cellular networks (Modems 1, 2, and 3) over a period of time. It is evident that no single network can offer latency shorter than 100ms during any of the considered time intervals}
    \label{fig:driveUdiagram-network}
\end{figure}

High latency is another crucial factor affecting the quality of teleoperation live video. Figure \ref{fig:driveUdiagram-network} shows the latency of our three cellular networks during a 4-second time period. While the transmission rate is almost constant and there is no congestion, clearly no single cellular network can offer sufficiently low latency, up to 100ms. Each cellular network has time periods during which its latency is above the 100ms horizontal dashed line, but, interestingly, these time periods are different for the different cellular networks. This suggests that sending different packets over different cellular networks can significantly improve video quality compared to sticking to the best single network.

We believe that machine learning (ML) is a good way to solve the difficult MPR problem, given the widespread use of AVs for traveling along predefined routes today. We conducted 100 communication test drives, each lasting approximately 15--20 minutes, in the same geographical area, using the same route. We then captured important network communication parameters, including GPS longitude, GPS latitude, RSRP, RSRQ, RSSI, bit rate, packet loss rate, bandwidth, and latency. We then use these parameters to train our ML algorithms.

We present three ML algorithms. The first algorithm is called LossPredict. This algorithm uses the above features to predict which of the cellular networks is more likely to deliver the next packet with no loss. The second proposed ML algorithm is called HandPredict. It uses the above features to predict which of the cellular networks is less likely to experience a handover while the next packets are transmitted. While both algorithms perform very well, with prediction accuracy of more than 80\%, we found HandPredict to be better in terms of the tradeoff between loss prediction and cost. This algorithm needs fewer parameters than LossPredict to achieve better packet loss prediction. The third proposed algorithm is called LatencyPredict. It uses the above features to predict the latency of the next video packets for each cellular network.

ML algorithms can detect invisible patterns in network behavior, such as deteriorating signal quality or decreasing bitrate. By leveraging these parameters, the proposed algorithms perform much better than, for example, a non-learning algorithm that relies only on GPS coordinates or on RSRP and RSRQ values for predicting network quality. Our work reveals that relying only on GPS coordinates, or only on RSRP and RSRQ values, to predict network quality is insufficient.

Our contribution is a generic approach for selecting the cellular network over which the next video packets will be transmitted by an AV. We defined a new problem, MPR, and propose a new algorithm, called ANS, to solve it. ANS aims to minimize the packet loss rate and packet latency with no prior knowledge of the geographical area, the AV mobility pattern, or the cellular network coverage. Commercial companies can use the proposed framework by modeling a geographical area with test drives, recording the relevant features, and training the proposed ML models. 

The rest of this paper is structured as follows: Section \ref{RELATEDSEC} reviews related work. Section \ref{methodology} describes the end-to-end forecasting framework of the three proposed ML algorithms. Section \ref{training-res} presents training results. Section \ref{eval-res} evaluates the trained ML algorithms on out-of-sample communication drives. Section \ref{ppNS} presents an algorithm that uses these ML algorithms and compares our results to a baseline algorithm used today in commercial systems. Finally, Section \ref{conclusion} concludes the paper.

%% file: sections/2-related-work.tex
Network communication prediction is not a new idea \cite{chiu2000predictive}. Many studies have been conducted to develop ways for efficient management of users in a cellular network. With the introduction of 5G networks, new efforts have been made to anticipate future cellular handovers and support various quality of service-sensitive \cite{abdah2020handover, ahmad2018efficient}. 

In \cite{abdah2020handover}, the authors rely on predictable user patterns. Since they assume a fixed path, the antennas of the serving cells' are known to the handover prediction. In contrast, in our work, we use data from communication test drives and learn from it without any prior assumptions. 

The authors of \cite{tan2022intelligent} address the problem of handover decision-making for moving vehicles across different base stations. They use deep reinforcement learning to learn an optimal handover policy that minimizes the number of unnecessary handovers and maximizes network throughput. While they consider a single cellular network, we consider multiple networks and define a different optimization problem. 

In \cite{https://doi.org/10.48550/arxiv.2111.13879}, the authors propose a data-driven ML scheme to solve the handover prediction task. They employ a multi-layer perceptron for the ML model based mainly on the received signal strength from the simulations. Again, their work is different from our work since they consider a single cellular network and a different optimization problem. Moreover, unlike our work, their results are based on ns-3 \cite{riley2010ns} simulations and not on real data.

The work of \cite{fattore2020automec} investigates user mobility prediction in automotive scenarios with the use of long short term memory (LSTM) recurrent neural network configurations. Similar to us, the authors show that LSTM can provide accurate mobility predictions. In contrast to our work, their results are based on simulations and not real data.

The authors of \cite{chen2013predicting} leverage channel state information coupled with the user's handover history for supervised ML that predicts future handovers. This approach relies on user equipment to report channel gain to the base station on a regular basis, which may delay or distort the forecast. In contrast, our ML predictors are trained offline, so the only inference of our ML trained models is the time it takes to calculate the prediction.

%% file: sections/4-methodology.tex
\begin{figure*}[!tb]
    \centering
    \includegraphics[width=0.9\textwidth]{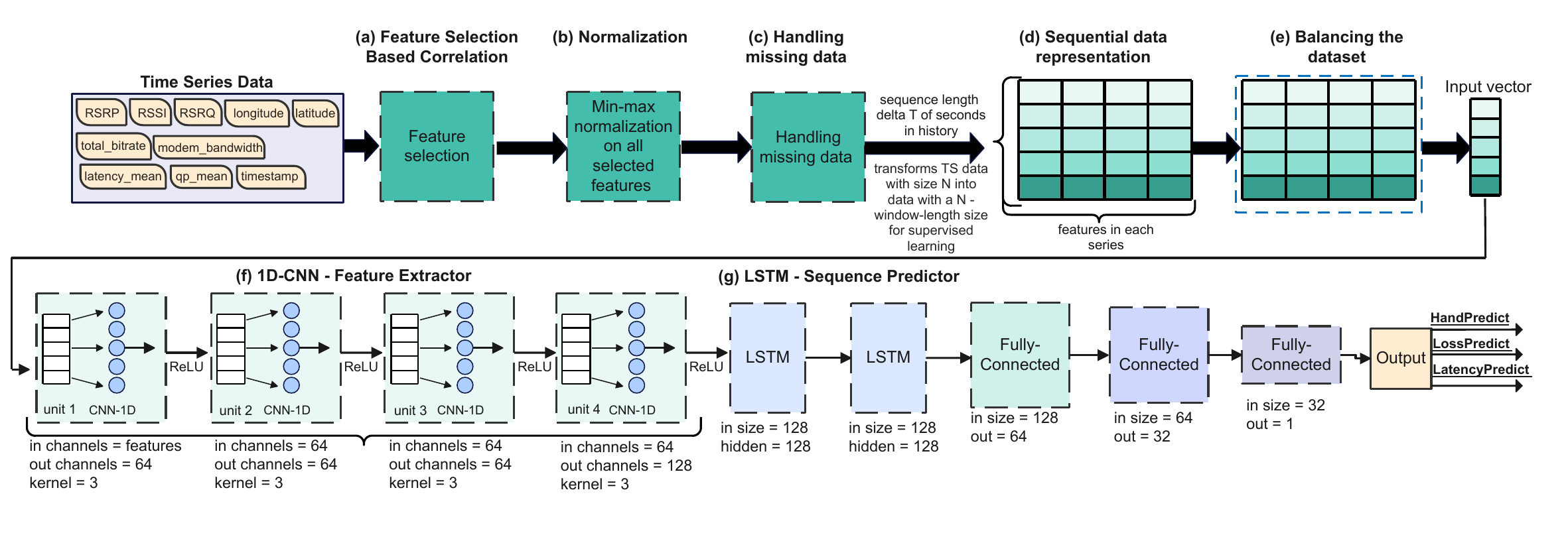}
    \caption{The pipeline stages in the proposed ML framework}
    \label{fig:framework}
\end{figure*}
In Section \ref{INTRO} we defined the MPR problem as a decision made by an AV connected to multiple cellular networks about over which cellular network to send the next video packets to ensure high-quality real-time video streaming. In this section, we transform the MPR problem into a time series forecasting problem. Time series data comprises a discrete or continuous set of values recorded over time. In the context of this work, each time record represents $1$ second and contains several parameters such as RSSI, RSRP, latency, GPS longitude, and GPS latitude. We then present the three ML algorithms: HandPredict for handover prediction, LossPredict for packet loss prediction, and LatencyPredict for latency prediction. The three algorithms share the same framework but have different ML loss functions and activation functions.

For HandPredict, we denote the predicted binary outcome at time $t+H$ as $\hat{Y}_{t+H}$, where $H$ represents the prediction horizon. This value is based on a probabilistic model with parameters $\theta$. The mathematical representation is as follows: 
\[{\Hat{Y}_{t+H}} = \begin{cases} 
\centering
  $1$ & \mbox{ if $P_\theta(Y_{t+H}=1|X_{t-T},..,X_{t}) \geq D_\mathrm{thresh}$}\\
  $0$ & \mbox{ if $P_\theta(Y_{t+H}=1|X_{t-T},..,X_{t}) < D_\mathrm{thresh},$}
\end{cases}\]
where $P_\theta(Y_{t+H}=1|X_{t-T},..,X_{t})$ is the conditional probability of $Y_{t+H}$ being equal to $1$ given the input features $X_{t-T},..,X_{t}$, $t$ is the current time step, and $T$ is the number of time steps included in the past window input features. This equation compares the calculated probability to a decision threshold $D_\mathrm{thresh}$. If the probability is greater than or equal to the threshold, the predicted outcome $\hat{Y}_{t+H}$ is set to $1$. Otherwise, it is set to $0$. For LossPredict and LatencyPredict, the ML models operate within a regression framework, thus predicting a continuous value with a $1$-second prediction horizon. Our aim is to meet the 100ms constraints. Thus, our predictions are made 1 second ahead of each step.

Figure \ref{fig:framework} presents the proposed ML framework for HandPredict, LossPredict, and LatencyPredict. This framework begins with five data preprocessing stages ((a)--(e) in the figure). The first preprocessing stage is feature selection, which determines the model input. Then, the input data passes through a normalization stage (b) and a missing data imputation stage (c). The time series data are then transformed into sequential data with input and output pairs (stages (d) and (e)). The model subsequently uses a 1D-CNN \cite{kiranyaz20211d} stage (f), which takes the sequential data and extracts features from it. Next, there is a two-LSTM layer stage (g), during which the LSTM \cite{yu2019review} learns nonlinear information derived from the output layers of the CNN. The two-LSTM layer uses guidance from patterns discovered by the earlier layers to detect important hidden information during every time period. Lastly, we use different ML loss and activation functions for each of the algorithms. Below, we explain each stage in more detail.
\begin{figure}[tb]
    \centering
    \includegraphics[width=0.7\columnwidth]{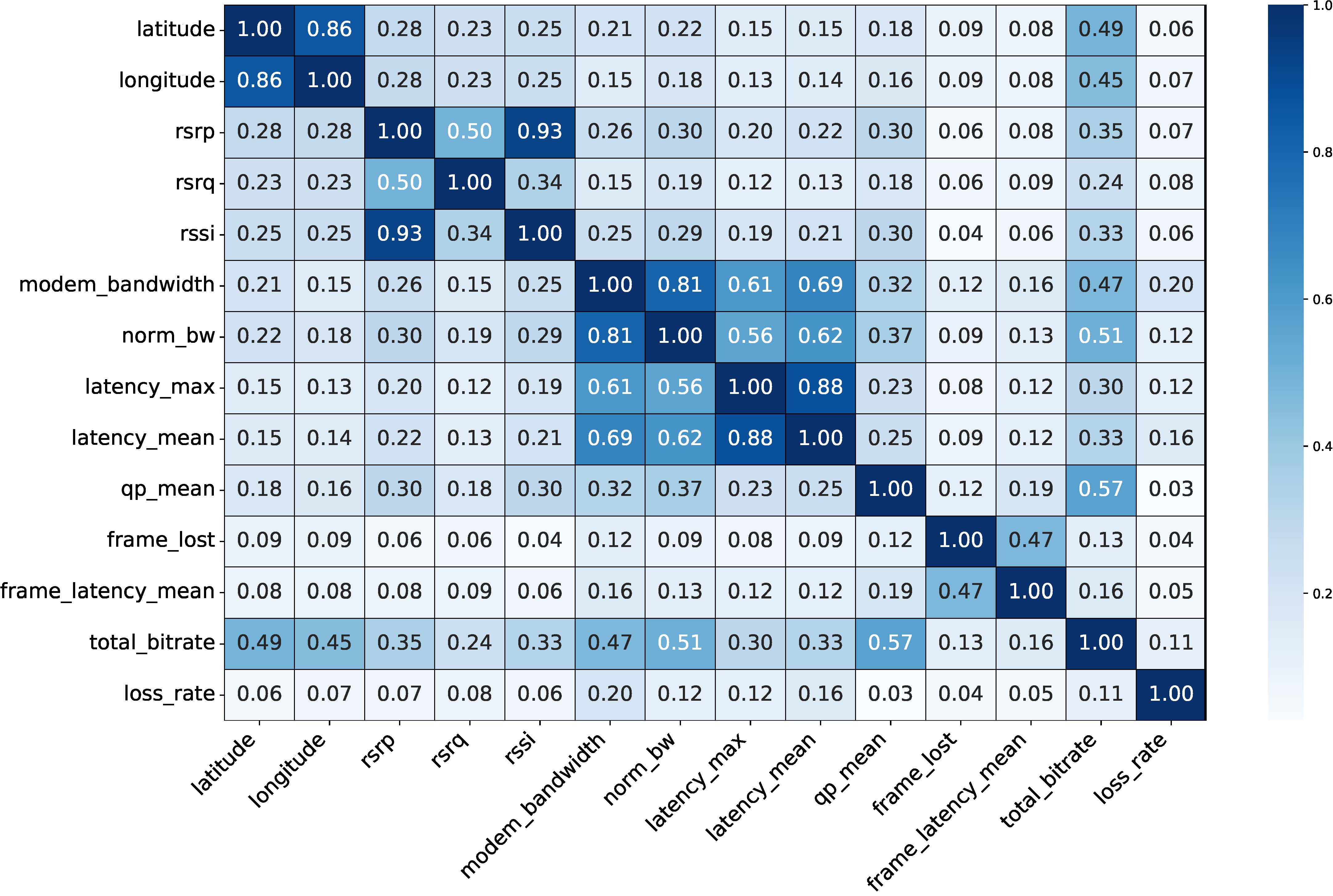}
    \caption{A correlation matrix that represents the relationships between pairs of features in a dataset of communication drives; each value in this matrix represents the strength of the relationship between the two corresponding features; 
    }
    \label{fig:CORR}
\end{figure}

\textbf{(a) Feature selection:} The selected features play a significant role in any ML model. In our problem, incorporating features that have a large impact on the handover probability or packet loss prediction enhances the model's accuracy. Increasing the number of features, however, also increases the input dimension of the CNN model, which leads to more parameters to learn. To address this, we employ a correlation analysis technique \cite{rehfeld2011comparison} that selects the most important features and removes those that have a high correlation with the selected features. Specifically, we remove features whose correlation coefficients with the selected features are $>$ 0.9. Figure \ref{fig:CORR} shows a correlation matrix between the various features in a set of test drives. It is evident from this matrix, for example, that RSRP, RSSI, and RSRQ are highly correlated with each other, which means that one of them can be omitted.

\textbf{(b) \textbf{Normalization}:} To use different variables with different units and size ranges, all variables are normalized to $[0,1]$. We use min--max normalization \cite{patro2015normalization}, which converts each pair of input and output variables, $X$ and $Y$, as follows: $x^\prime_{ij} = \frac{x_{ij}-x^\mathrm{min}_{j}}{x^\mathrm{max}_{j}-x^\mathrm{min}_{j}}, \,\,\, 
$$y^\prime_{i} = \frac{y_{ij}-y^\mathrm{min}}{y^\mathrm{max}-y^\mathrm{min}}.$ Here, $x_{ij}$ and $x^{\prime}_{ij}$ are the original and normalized values of the $j$$^\mathrm{th}$ input variable in the $i$$^\mathrm{th}$ input sample, respectively; $y_i$ and $y^{\prime}_i$ are the original and normalized values in the $i$$^\mathrm{th}$ output sample, respectively; $x^\mathrm{min}_{j}$ and $x^\mathrm{max}_{j}$ are the minimum and maximum values of the j$^\mathrm{th}$ input variable in all i$^\mathrm{th}$ input samples, respectively; finally, $y^\mathrm{min}$ and $y^\mathrm{max}$ are the minimum and maximum values of all output samples, respectively.

\textbf{(c) Handling missing data:} Real datasets are usually incomplete, with many missing values. For example, in our dataset, 36\% of the points lack their ``GPS longitude" and ``GPS latitude" features. Missing dataset values are a challenge for most ML techniques. To address this challenge, a common practice is to identify and replace missing values before modeling the prediction task, a technique known as missing data imputation \cite{ahn2022comparison, https://doi.org/10.48550/arxiv.2103.01600}. A popular algorithm for missing data imputation is the $k$-nearest neighbor (KNN) \cite{peterson2009k}. Using KNN for each sample with missing values, the algorithm finds $k$ closest samples in the dataset based on Euclidean distance and uses their mean instead of the missing value. We use this method with $k=2$.

\textbf{(d) Sequential data representation:} Our dataset consists of time series data obtained by DriveU Ltd., which covers several cellular networks, and includes multiple features such as RSRQ, RSRP, GPS latitude, GPS longitude, etc. We represent the time series data as a matrix, with rows representing 1-second time steps and columns representing various features. We use 1-second time steps in order to meet the strict video end-to-end latency requirements.

To apply supervised learning to time series data, the data must be converted into input--output pairs. To this end, we use the sliding window technique \cite{frank2001time}, which requires two parameters: the sliding window length and the sliding step. We set both to $1$ second, for predicting ten time steps ahead. We use sliding window lengths of $32$, $64$, and $128$ seconds, resulting in $32$, $64$, and $128$ rows of multiple features for each training sample, respectively. Since the sliding step is $1$ second long, there is an overlap between the windows and the maximum number of training samples is obtained. 

In supervised learning, classification tasks and regression tasks require labeled data. For handover prediction, we identify two significant events as handover labels: the label is $1$ if the serving cell changes from X to Y in consecutive time steps, and $0$ if there is no change. For the packet loss rate prediction task, the packet loss rate recorded data on time step $t+1$ is used as the label for each sample.

\textbf{(e) Balancing the data:} An imbalanced dataset can be biased towards the majority class. This bias can lead to poor performance among the minority population and decrease overall accuracy. By balancing the dataset, the ML model is trained on a representative sample of the entire dataset, thereby making it more general. We use an undersample technique \cite{yen2006under}, which removes a portion of the data from the majority class, to balance the dataset for the handover prediction. Undersampling does not affect the correlation structure within each window in the original sequential data. To this end, we count the number of samples in the dataset whose handover label is $1$, and find an imbalance ratio of $2:98$, indicating that only $2\%$ of the samples have handovers. To ensure a balanced training set with a $50:50$ ratio, we randomly choose the same number of samples with and without handovers from the same test drives. For packet loss prediction, which is a regression task, we predict a continuous value and do not balance the dataset. Thus we can still use all the samples.

The last two stages, (f) and (g), contain the proposed neural network architecture, which is inspired by \cite{ordonez2016deep}. This includes using a CNN for feature extraction and incorporating LSTM layers, followed by fully-connected layers, for prediction.

\textbf{(f) Using a CNN for feature extraction:} It is common to use CNNs in image processing applications. Unlike convolutional CNNs, which use squared filters for image processing, the 1D-CNN uses rectangular filters to recognize features in a time series. In each rectangular filter, \textit{h} is the number of samples in a test drive, and \textit{w} is the number of features selected from the input data. A 1D-CNN can extract important temporal features using convolutional layers. To increase the nonlinear features of the CNN, we apply a ReLU activation function, which enhances the expression ability of the network. Typically, to minimize input complexity and ensure translation in-variance, a CNN framework involves successive convolutional and pooling layers. However, since the recurrent layers need to process a sliding window of temporal data, we exclude pooling operations from our architecture. We use four convolutional layers, each with three single-stride kernels.

\textbf{(g) LSTM for prediction:} An LSTM architecture \cite{8141873} is used for making the final prediction. First, the output of the CNN feature extractor is passed as input to the first LSTM layer. The output of the first LSTM layer is the input to the second LSTM layer, and then we have three fully connected layers that calculate the final prediction. 

Before choosing the above architecture, we tested several alternatives, such as: (a) a bi-directional LSTM with self-attention, which has higher complexity, but did not provide a better prediction; (b) a temporal convolutional network whose results were worse than those of our architecture.

For HandPredict, the output layer estimates the likelihood that a handover event will occur, which is subsequently compared to the actual label. The error value is computed using a binary cross-entropy ML loss function. For LossPredict and LatencyPredict, the output layer predicts a continuous value of the packet loss rate or latency, respectively. The error value is then computed using a mean square error and a mean absolute error (MAE), for the ML loss functions. Using these algorithms, an AV can anticipate a deteriorating signal quality or decreasing bitrate in the cellular network, and thus can determine in real-time over which cellular network the next video packets should be transmitted.

Figures \ref{fig:drive} and \ref{fig:driveUdiagram-network} demonstrate that different cellular networks behave differently in the same geographical area. For this reason, we used data from each cellular network to train separate instances of the same ML models. During training, after each epoch, the gradients of each layer in all the models were averaged. This ensures that each model contributes equally to the learning process, and that the model will perform consistently regardless of the underlying cellular network \cite{chen2018gradnorm}. During inference, the unified model uses the shared weights to predict the latency and the probability for a handover. This approach simplifies the deployment process since only one ML model has to be maintained. 

%% file: sections/5-training-results.tex
\begin{table}[!t]
\centering
\resizebox{1\columnwidth}{!}{%
\begin{tabular}{c|c|c|c|c|c}
\textbf{Model name}                 & \textbf{GPS-only} & \textbf{RSRP\textbackslash{}RSRQ} & \textbf{7-feature} & \textbf{8-feature} & \textbf{9-feature} \\ \hline \hline
\textbf{Features} &
  \begin{tabular}[c]{@{}c@{}}GPS longitude\\ GPS latitude\end{tabular} &
  \begin{tabular}[c]{@{}c@{}}RSRP\\ RSRQ\end{tabular} &
  \begin{tabular}[c]{@{}c@{}}Time stamp\\ RSRP\\ RSRQ\\ Modem bandwidth\\ Normalized bandwidth\\ Packet loss rate\\ Total bit-rate\end{tabular} &
  \begin{tabular}[c]{@{}c@{}}Time stamp\\ RSRP\\ RSRQ\\ Modem bandwidth\\ Normalized bandwidth\\ Total bit-rate\\ GPS longitude\\ GPS latitude\end{tabular} &
  \begin{tabular}[c]{@{}c@{}}Time stamp\\ RSRP\\ RSRQ\\ Modem bandwidth\\ Normalized bandwidth\\ Packet loss rate\\ Total bit-rate\\ GPS longitude\\ GPS latitude\end{tabular} \\ \hline
\multirow{3}{*}{\textbf{Algorithm}} & HandPredict       & HandPredict                       & HandPredict        &                    & HandPredict        \\ \cline{2-6} 
                                    &                   &                                   &                    & LossPredict        &                    \\ \cline{2-6} 
                                    & LatencyPredict    & LatencyPredict                    & LatencyPredict     &                    & LatencyPredict     \\ \hline
\end{tabular}%
}
\caption{The different sets of input features for each algorithm}
\label{tab:features-table}
\end{table}
\begin{figure}[t]
\centering
\captionsetup{justification=centering}
\begin{subfigure}{.49\columnwidth}
  \centering
  \captionsetup{justification=centering}
  \includegraphics[width=1\textwidth]{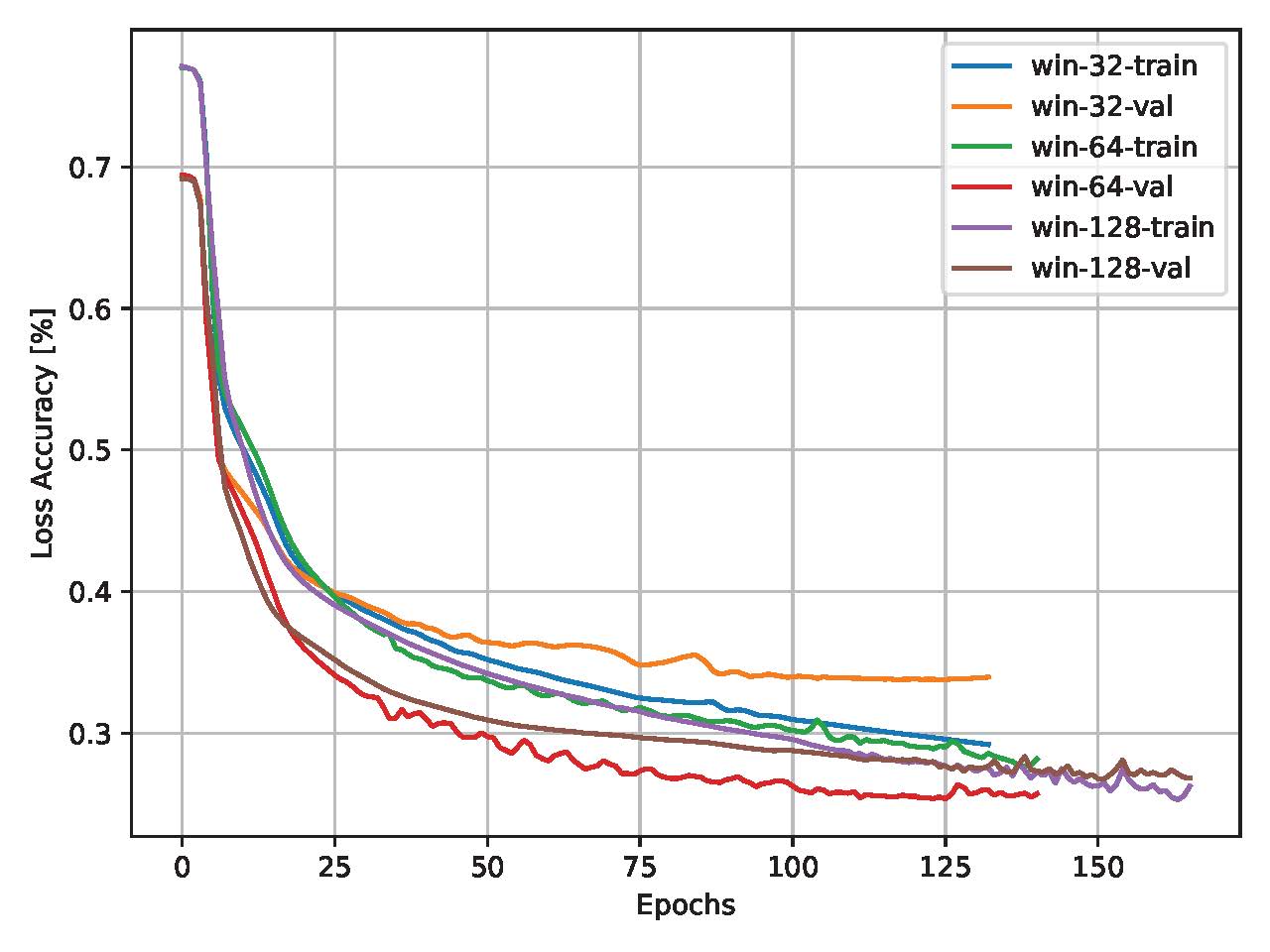}
  \caption{ML loss accuracy as a function of the time for different window lengths}
  \label{fig:trainloss-80}
\end{subfigure}
\begin{subfigure}{.49\columnwidth}
  \centering
  \captionsetup{justification=centering}
  \includegraphics[width=1\textwidth]{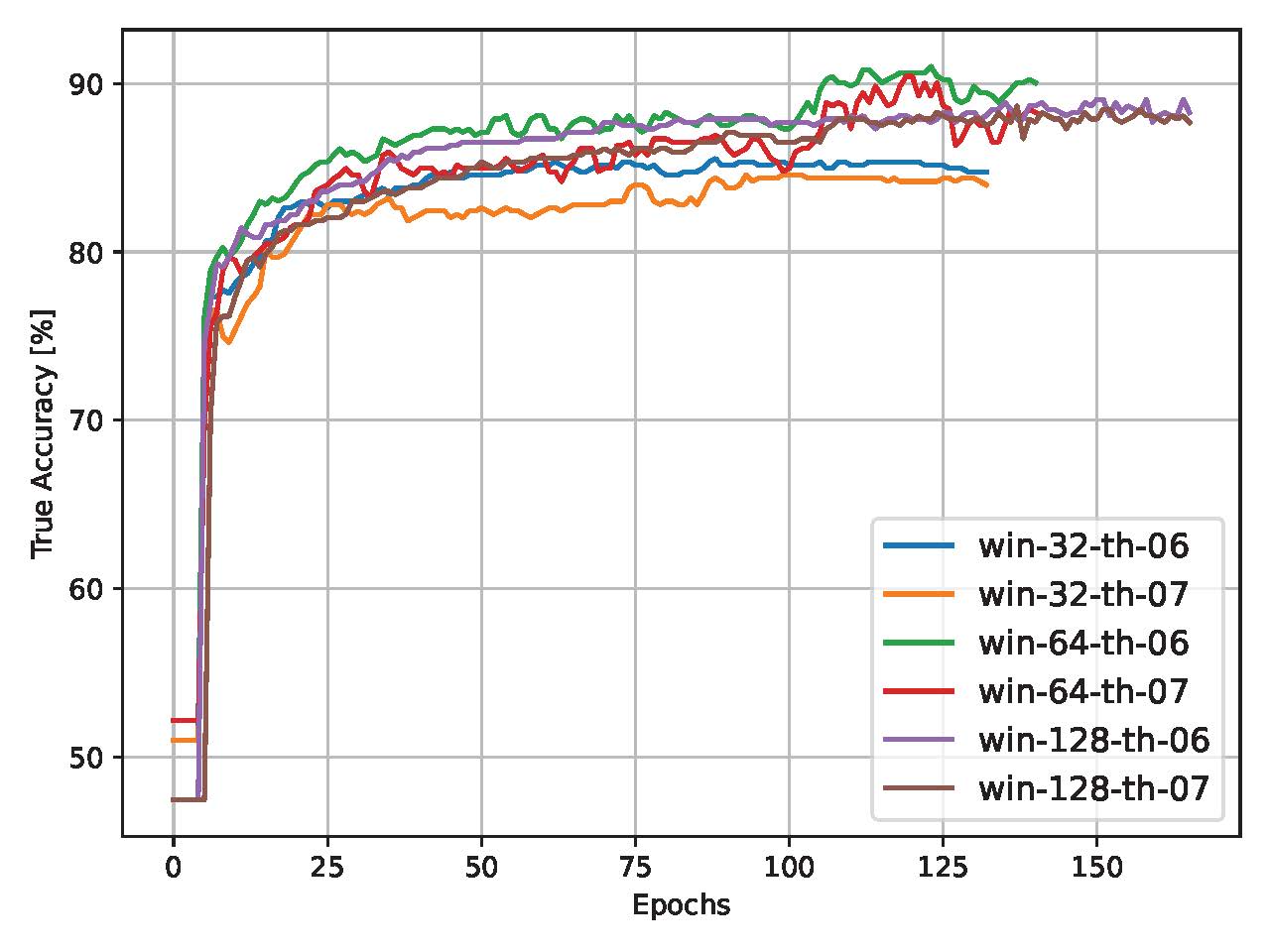}
  \caption{True accuracy as a function of the time for different window lengths and thresholds}
  \label{fig:trainacc-80}
\end{subfigure}%
\caption{The prediction dataset of HandPredict for 80 test drives (first configuration)}
\label{fig:training-80}
\end{figure}
\begin{figure}[!ht]
\centering
\captionsetup{justification=centering}
\begin{subfigure}{.49\columnwidth}
  \centering
  \captionsetup{justification=centering}
  \includegraphics[width=1\textwidth]{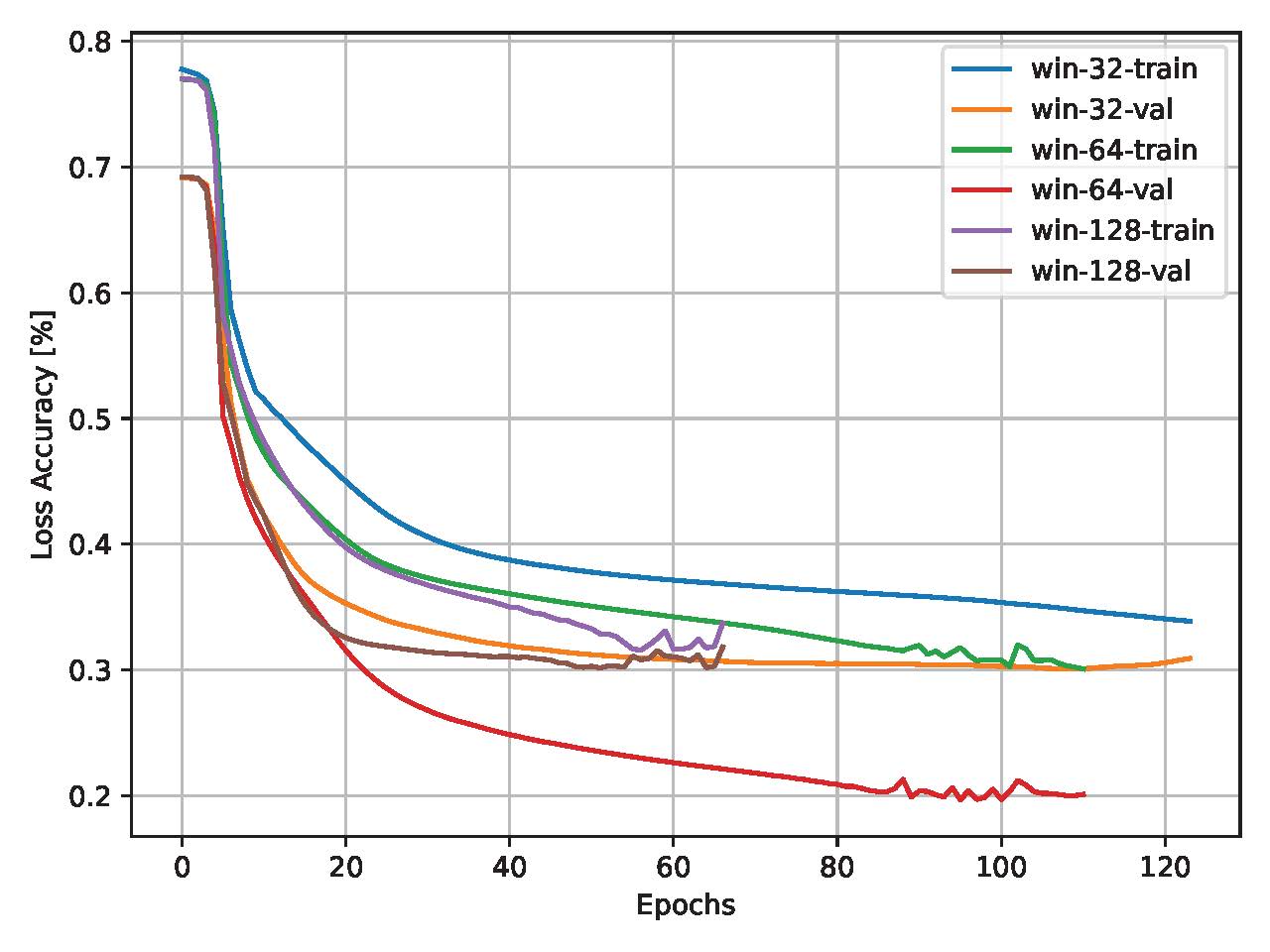}
  \caption{ML loss accuracy as a function of the time for different window lengths}
  \label{fig:trainloss-20}
\end{subfigure}
\begin{subfigure}{.49\columnwidth}
  \centering
  \captionsetup{justification=centering}
  \includegraphics[width=1\textwidth]{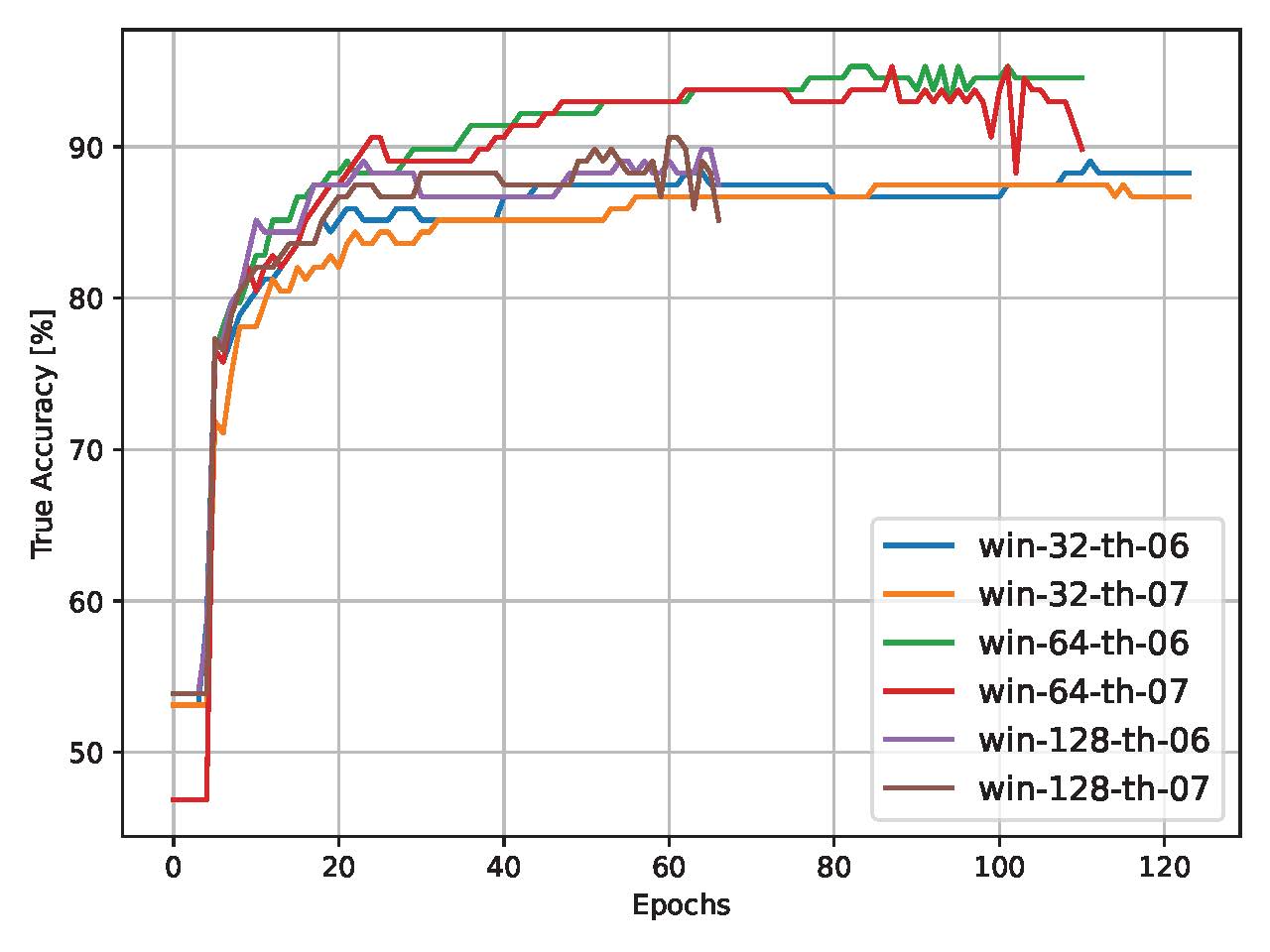}
  \caption{True accuracy as a function of the time for different window lengths and thresholds}
  \label{fig:accloss-20}
\end{subfigure}%
\caption{The prediction dataset of HandPredict for 20 test drives (second configuration)}
\label{fig:training-20}
\end{figure}

In this section, we show training results for HandPredict, LossPredict, and LatencyPredict. The results we show is for the unified neural network for each ML model. Table \ref{tab:features-table} presents the different sets of input features for each algorithm. The table shows the four HandPredict models and the four LatencyPredict models. During training, we found that predicting the probability of a handover event requires less training than predicting packet latency and packet loss rate. Therefore, HandPredict uses two separate datasets: a dataset with 80 test drives and a dataset with only 20 test drives. In contrast, LatencyPredict and LossPredict use only the large dataset, with 80 test drives. These drives conducted over a fixed route. Additionally, while we use four different models for HandPredict and LatencyPredict, only the 8-feature model is used for LossPredict. Without these features, predicting the packet loss rate is less accurate.

As Table \ref{tab:features-table} shows, one of our models uses only RSRP and RSRQ. The justification for this model is that the key to performing a handover process is by considering the RSRP and RSRQ measurements in the following way: The user measures the serving cell's RSRP and RSRQ, and compares these values to those of the neighboring cells. If the RSRP of the serving cell is below some threshold (often between -115 and -100 dBm), or if the RSRQ is below another threshold (often between -19.5 and -3 dB), or if the RSRP and RSRQ from the neighbouring cell are significantly better, a handover request is submitted.

To train the various ML models, the architecture of Figure \ref{fig:framework} is used, with two different configurations for stages (f) and (g). In the first configuration, the model is trained using a dataset of 80 test drives, each lasting 15--20 minutes. In this configuration, the batch size is set to 512, the learning rate to 0.001, the ML loss function is binary cross-entropy, and the activation function is sigmoid. In the second configuration, we train the model on a smaller dataset consisting of only 20 drives. In this case, the batch size is 128, and the learning rate is 0.001. 

The first set of results we show is for the HandPredict algorithm. Figure \ref{fig:training-80}(\subref{fig:trainloss-80}) shows the ML loss accuracy as a function of the sliding window length for the first configuration on both training and validation sets. Training is stopped when the ML loss accuracy on the validation set does not improve during several consecutive epochs. The considered window length varies from $32$ to $128$ seconds. It is evident that a 64-second sliding window yields the best performance. Figure \ref{fig:training-80}(\subref{fig:trainacc-80}) shows true accuracy as a function of sliding window lengths. The true accuracy metric is the ratio of accurate predictions made on the validation set. If the ML model predicted value is higher than $D_\mathrm{thresh}$, the prediction is considered a handover event. Figure \ref{fig:training-80}(\subref{fig:trainacc-80}) shows the true accuracy for two different $D_\mathrm{thresh}$: $0.6$ and $0.7$. It is evident that a sliding window length of 64 yields the best true accuracy. Figure \ref{fig:training-20}(\subref{fig:trainloss-20}) and Figure \ref{fig:training-20}(\subref{fig:accloss-20}) are similar to Figure \ref{fig:training-80}(\subref{fig:trainloss-80}) and Figure \ref{fig:training-80}(\subref{fig:trainacc-80}), respectively, but use the second configuration.

\begin{figure}[!t]
  \centering
  \captionsetup{justification=centering}
  \includegraphics[width=0.7\columnwidth]{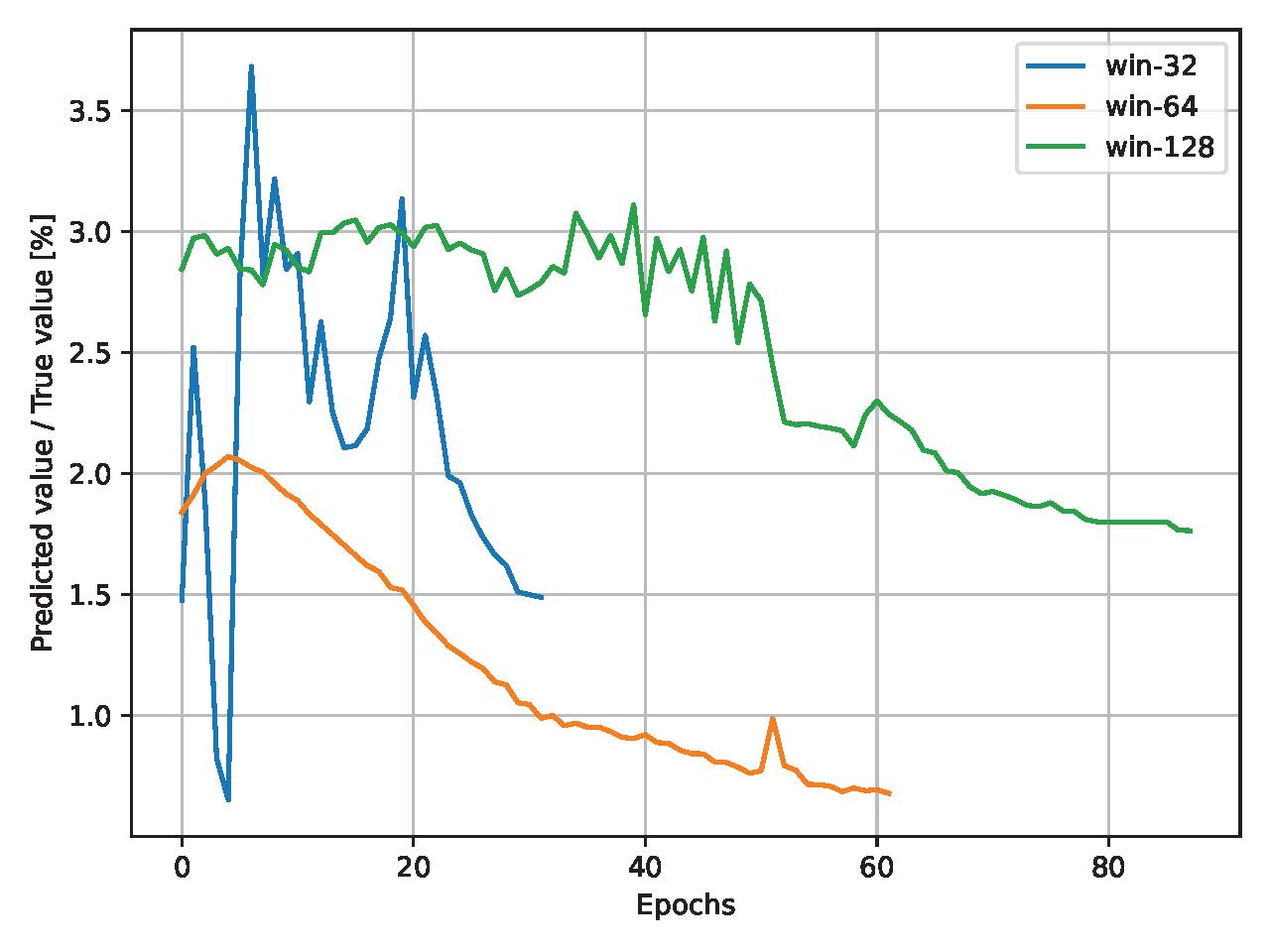}
  \caption{Packet loss prediction ratio of LossPredict on a validation dataset of 80 test drives}
  \label{fig:loss-pred:training}
\end{figure}%

The second set of results is for the LossPredict algorithm. For this model, only one configuration is used. The model is trained using a dataset of 80 test drives, each lasting 15--20 minutes. The batch size is set to 512, the learning rate to 0.001, the ML loss function is mean square error, and the activation function is the identity function.

Figure \ref{fig:loss-pred:training} shows the calculated percentage of correct prediction of training on the validation set. The x-axis indicates the epoch number during training. The y-axis represents the ratio between the predicted value and the real value. In this graph, ``1" on the y-axis indicates that the predicted value is equal to the real value of the packet loss rate when checked on the validation set. Recall that the prediction of a packet loss rate is a continuous value.

It is evident that for LossPredict, the 32-second sliding window outperforms the 128-second counterpart, as its accuracy ratio falls below 2; yet, it does not achieve the accuracy of the 64-second window, probably due to insufficient feature capture. For the 64-window length, the LossPredict algorithm achieves approximately 80\% accuracy around the 50$^\mathrm{th}$ epoch when the ratio is near 1. The ML model's accuracy, using a 128-second window, shows minor improvement after 80 training epochs, possibly due to the size of the neural network requiring more parameters for learning. This suggests that future ML models could benefit from increasing the number of layers and neurons.
\begin{figure}[!ht]
\centering
\captionsetup{justification=centering}
\begin{subfigure}{.50\columnwidth}
  \centering
  \captionsetup{justification=centering}
  \includegraphics[width=1\textwidth]{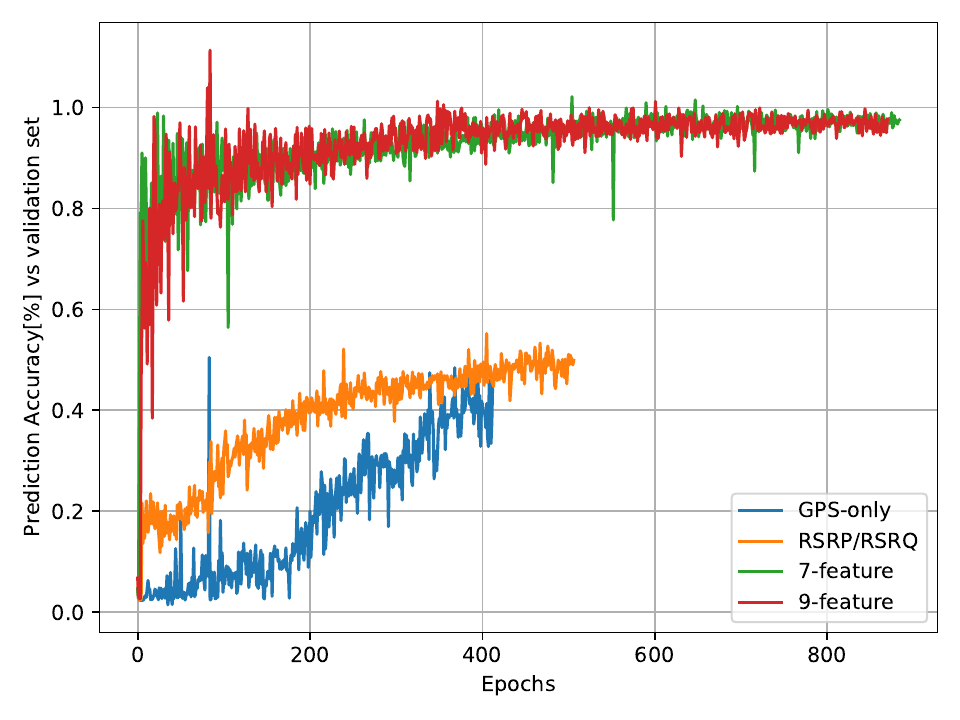}
  \caption{32 window length}
  \label{fig:latency-32-training}
\end{subfigure}
\begin{subfigure}{.50\columnwidth}
  \centering
  \captionsetup{justification=centering}
  \includegraphics[width=1\textwidth]{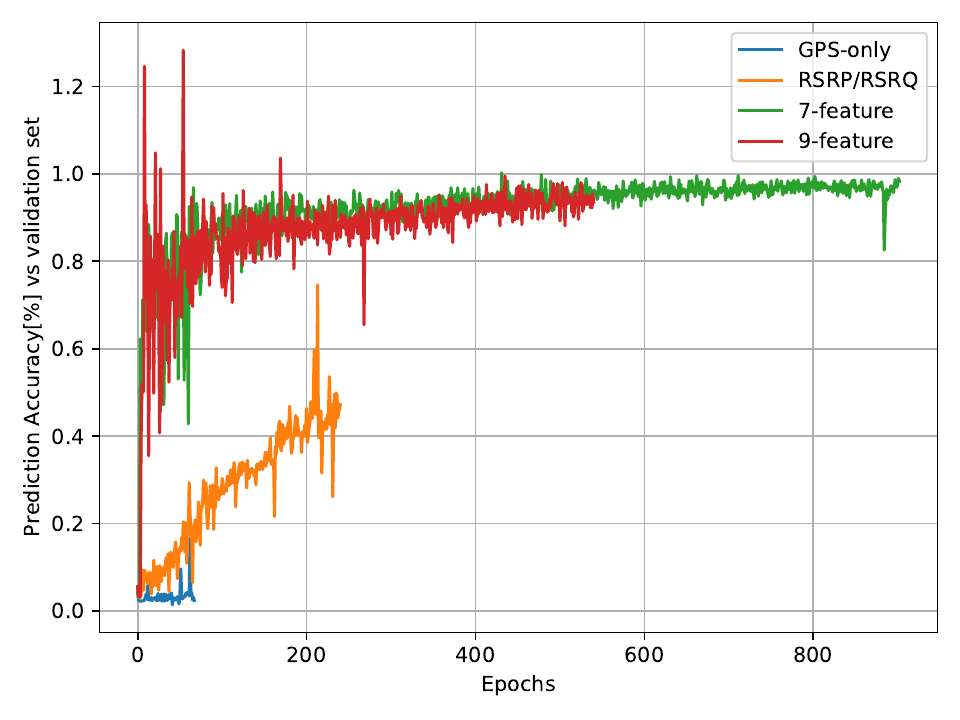}
  \caption{64 window length}
  \label{fig:latency-64-training}
\end{subfigure}%
\begin{subfigure}{.50\columnwidth}
  \centering
  \captionsetup{justification=centering}
  \includegraphics[width=1\textwidth]{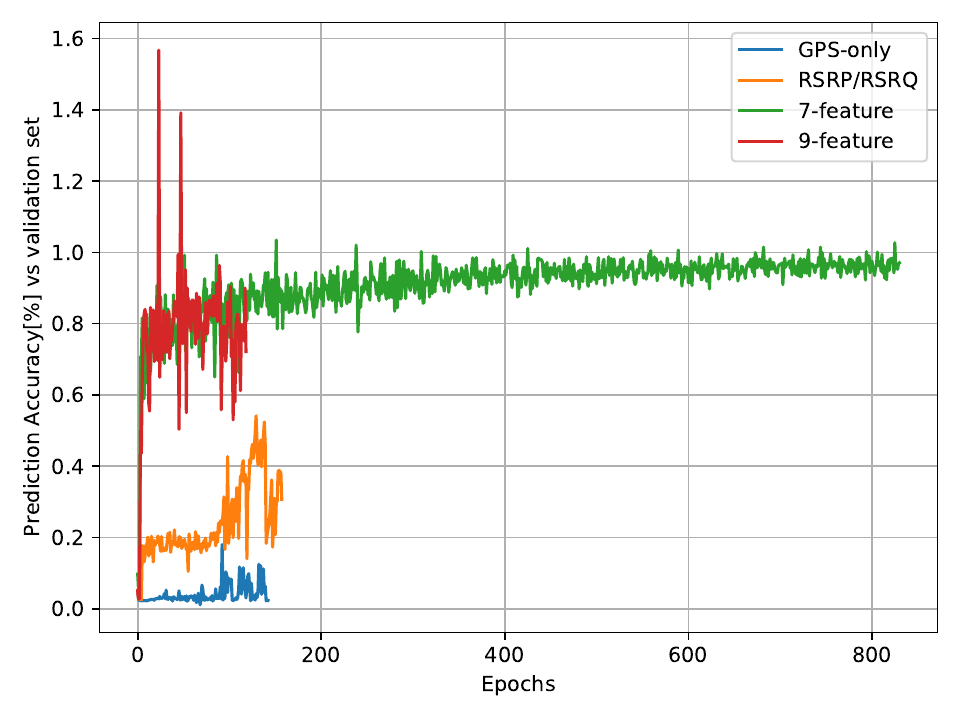}
  \caption{128 window length}
  \label{fig:latency-128-training}
\end{subfigure}%
\caption{Latency prediction ratio of LatencyPredict on a validation dataset of 80 test drives}
\label{fig:latency-training}
\end{figure} 

The third set of results is for the LatencyPredict algorithm. This algorithm has the same four models as HandPredict. These models are trained using a dataset of 80 test drives, each lasting 15--20 minutes. The batch size is set to 512, the learning rate to 0.001, the ML loss function is the MAE, and the activation function is the identity function.

Figure \ref{fig:latency-training} shows the calculated percentage of the correct latency prediction on the validation set. The x-axis indicates the epoch number during training and the y-axis represents the ratio between the predicted and real latency values. Thus, the value ``1" on the y-axis indicates that the predicted value is equal to the real value of the latency when checked on the validation set. Recall that the packet latency prediction is a continuous value.

It is evident that for LatencyPredict, the 7-feature and 9-feature models achieve the best prediction on the validation set, regardless of their window length. On the other hand, the GPS-only and RSRP/RSRQ models achieve low accuracy on the validation set. This indicates that using only these features might not be enough for latency prediction.

%% file: sections/6-evaluation-results.tex
\begin{figure}[!tb]
\centering
\captionsetup{justification=centering}
\begin{subfigure}{1\columnwidth}
\centering
  \captionsetup{justification=centering}
  \includegraphics[width=0.7\columnwidth]{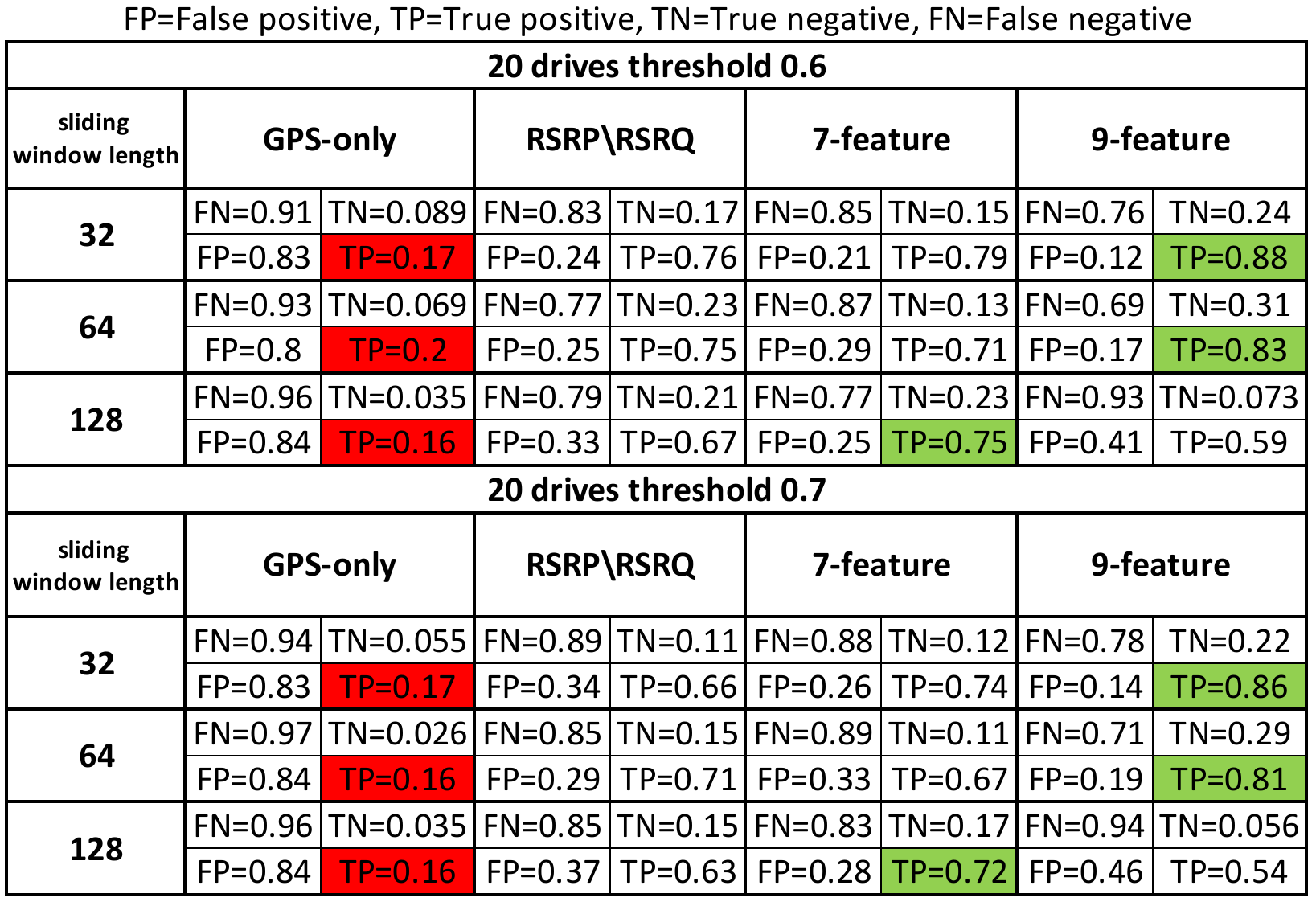}
  \caption{Dataset of 20 test drives}
  \label{fig:results-20-drives-confusion}
\end{subfigure}
\begin{subfigure}{1\columnwidth}
  \centering
  \captionsetup{justification=centering}
  \includegraphics[width=0.7\columnwidth]{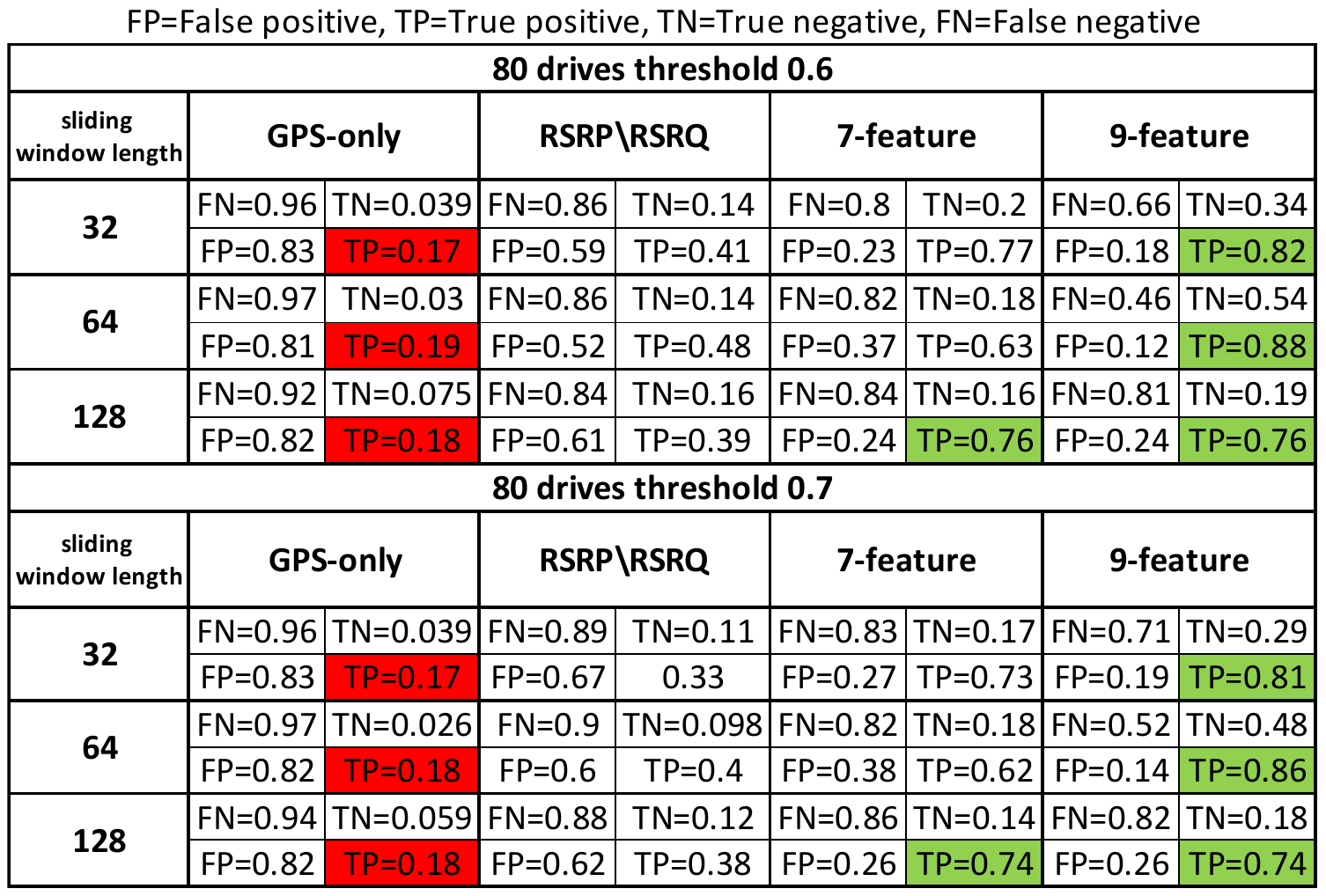}
  \caption{Dataset of 80 test drives}
  \label{fig:results80-drives-confusion}
\end{subfigure}
\caption{A confusion matrix for each of the four models of HandPredict with different sliding window lengths}
\label{fig:results-features}
\end{figure}
This section presents evaluation results from $20$ test drives that were not used during training. Figure \ref{fig:results-features} presents confusion matrices for HandPredict's four different models (see Table \ref{tab:features-table}): the 9-feature model, which uses all nine features, the 7-feature model, which uses a reduced set of input features, the GPS-only model, and the RSRP/RSRQ model. The figure highlights the best and worst true positive (TP) accuracies in green and red, respectively, from which we draw the following conclusions:
\begin{figure}[tb]
\centering
\captionsetup{justification=centering}
\begin{subfigure}{0.8\columnwidth}
  \centering
  \captionsetup{justification=centering}
  \includegraphics[width=0.7\textwidth]{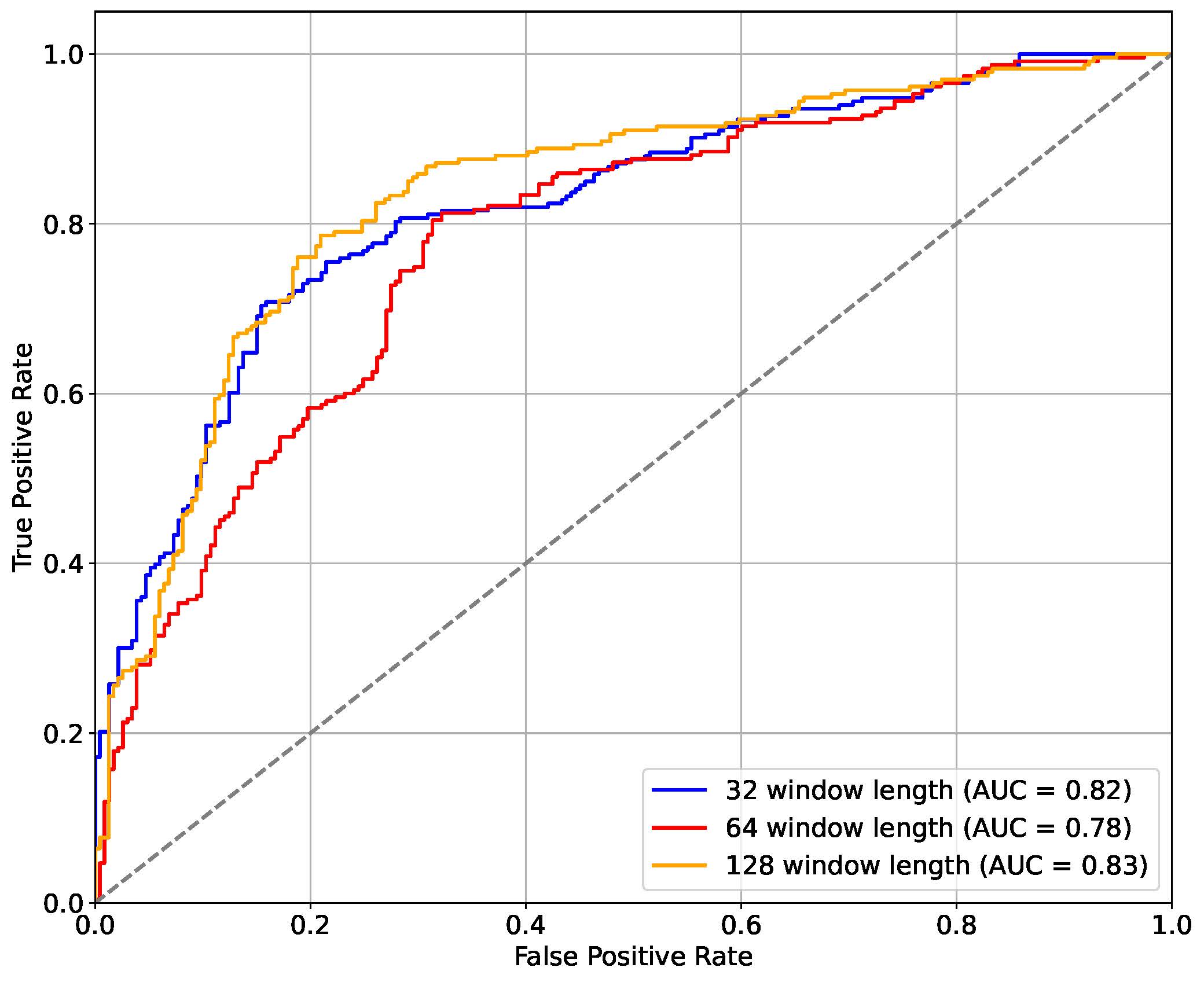}
  \caption{The 9-feature model with different window lengths (32,64,128)}
  \label{fig:diff-sliding-window-all-features}
\end{subfigure}
\begin{subfigure}{0.8\columnwidth}
  \centering
  \captionsetup{justification=centering}
  \includegraphics[width=0.7\textwidth]{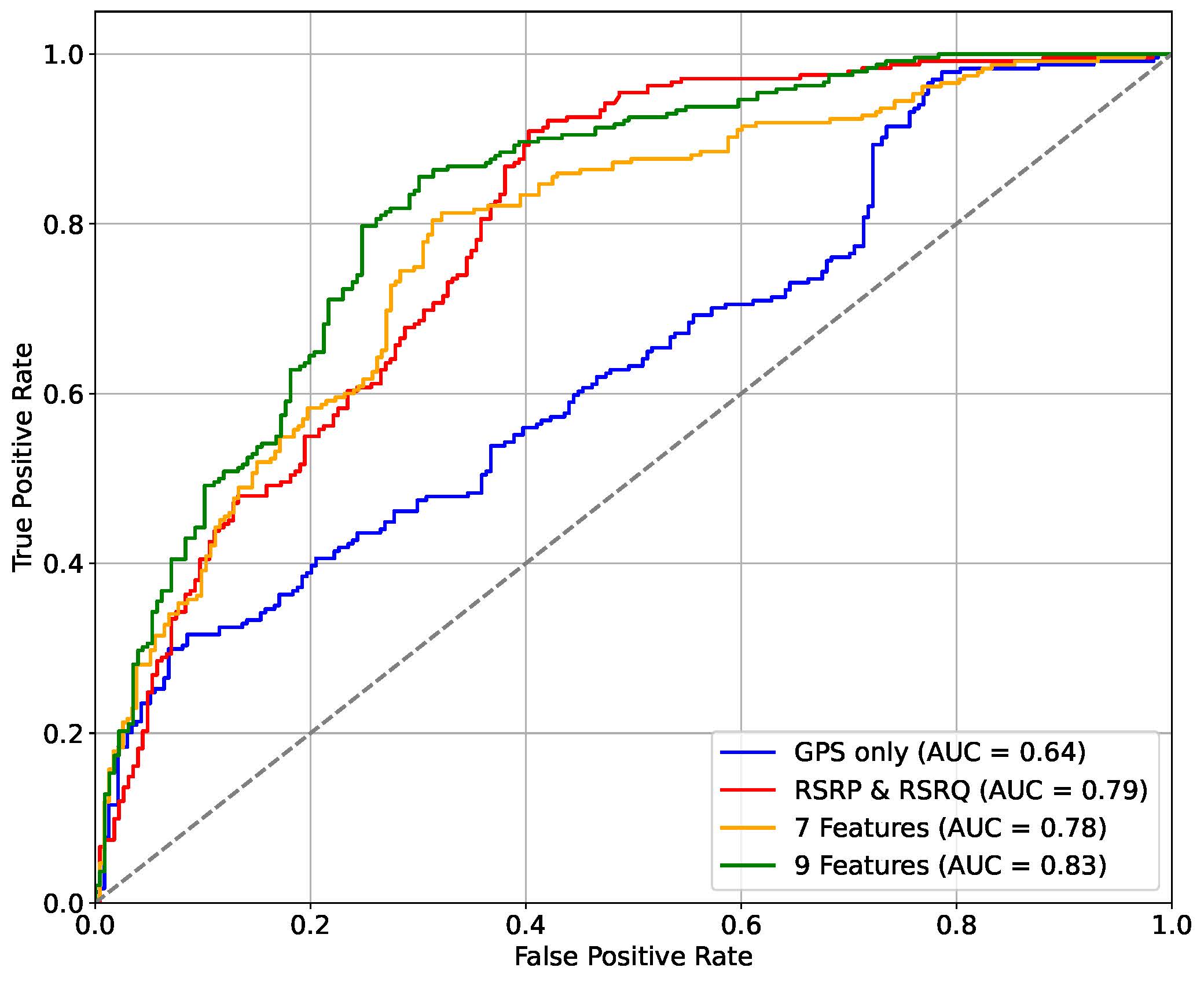}
  \caption{A 64-second window with four different models of HandPredict}
  \label{fig:64-window-diff-features}
\end{subfigure}%
\caption{TP vs. FP of HandPredict for several combinations of input features and window lengths}
\label{fig:ROC}
\end{figure} 
\newline $\bullet$ GPS coordinates alone are insufficient for predicting handover events. Using only RSRP and RSRQ yields better performance but is also not enough.
\newline $\bullet$ For the larger dataset, consisting of 80 test drives, the 128-second window length has the worst TP accuracy when all input features are used. A 64-second window length brings the highest TP accuracy of 88\%, while a 32-second window length achieves a 86\% TP accuracy. These results correspond to a strict threshold of 0.7. For the 7-feature model, 32- and 128-second windows have a similar confusion matrix, while a 64-second window has a 10\% better TP accuracy. The GPS-only model and the RSRP/RSRQ-model have low TP accuracy.
\newline $\bullet$ The number of input features used for each ML model has a significant impact on the prediction accuracy, and in particular on the TP accuracy. The GPS-only model has a very low TP. The RSRP/RSRQ-model has a better TP accuracy than the GPS-only model but it is worse than the 7-feature model. The 9-feature model achieves the best TP accuracy -- $88\%$.
\newline $\bullet$ Longer windows do not necessarily improve TP accuracy.

The receiver operating characteristic (ROC) curve illustrates the trade-off between the TP and false positive (FP) values of an ML model. Figure \ref{fig:ROC} shows the ROC curve for different models and window lengths. The models whose curves are closer to the top-left corner have better performance than those whose curves are closer to the bottom-right corner. The curve of an ML model whose decisions are made randomly is expected to be close to the diagonal line, which represents $FP=TP$. The figure also indicates the area under the ROC curve (AUC) for each model. The AUC range is $[0,1]$. For ML models whose predictions are 100\% accurate, AUC is 1, whereas for those whose predictions are 0\% accurate, AUC is 0.

Figure \ref{fig:ROC}(\subref{fig:diff-sliding-window-all-features}) shows the trade-off between TP and FP for the 9-feature model of HandPredict when trained with 3 different sliding window lengths: 32, 64, and 128 seconds. The results show that a 128-second window yields the best trade-off, as its curve is closest to the top left corner. With respect to TP accuracy only, a 64-second window brings the best results. Figure \ref{fig:ROC}(\subref{fig:64-window-diff-features}) compares the four different HandPredict models trained with a 64-second window. The figure shows that the 9-feature model achieves the best trade-off between TP and FP with an AUC of 0.83. The 7-feature model and the RSRP/RSRQ model exhibit comparable results, with AUCs of 0.79 and 0.78 respectively. The GPS-only model has the worst results.
\begin{table}[!tb]
\centering
\footnotesize
\begin{tabular}{c|c}
\textbf{Window Length} & \textbf{Accuracy [\%]} \\ \hline \hline
32 & 65  \\ \hline
64 & 80  \\ \hline
128 & 61 \\
\end{tabular}%

\caption{Packet loss prediction accuracy of LossPredict for the 8-feature model with unseen test drives, using a dataset of 80 test drives}
\label{tab:resLOSS}
\end{table}

Table \ref{tab:resLOSS} shows the packet loss rate prediction accuracy for unseen test drives, using LossPredict with the 8-feature model, for a dataset of 80 test drives. The considered window lengths are 32, 64, and 128 seconds. It is evident  that a 64-second window yields the best prediction accuracy. The 32-second window and the 128-second window require refinement of more parameters, such as the learning rate and the number of trainable parameters, to achieve good performance.

Evaluating the tradeoff between prediction accuracy and cost, it is evident that HandPredict outperforms LossPredict. Table \ref{tab:resLOSS} and Figure \ref{fig:results-features} show that HandPredict has better TP accuracy across various configurations. When compared against LossPredict, HandPredict is usually comparable or better. Furthermore, HandPredict is more efficient as it can also use a dataset of 20 test drives, in contrast to LossPredict, which must use larger datasets. This implies that LossPredict is more expensive, both in terms of the number of features it uses and the number of test drives required for training.

\begin{figure}[t]
  \centering
  \captionsetup{justification=centering}
  \includegraphics[width=0.7\columnwidth]{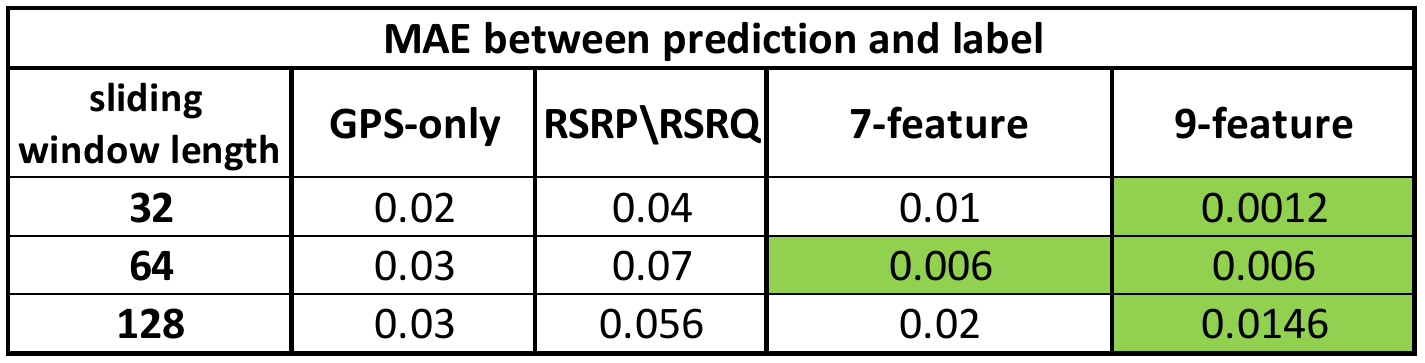}
  \caption{Latency prediction accuracy for each of the four LatencyPredict models, with different window lengths }
  \label{fig:latency-eval}
\end{figure}

Figure \ref{fig:latency-eval} shows the MAE latency prediction accuracy for LatencyPredict's four models on new test drives. It is evident that the 9-feature model achieves the best accuracy. It is also clear that the length of the sliding window does not significantly affect the results in this case.

%% file: sections/8-ppNS.tex
\begin{figure}[!tb]
    \centering
  \captionsetup{justification=centering}
  \includegraphics[width=0.8\linewidth]{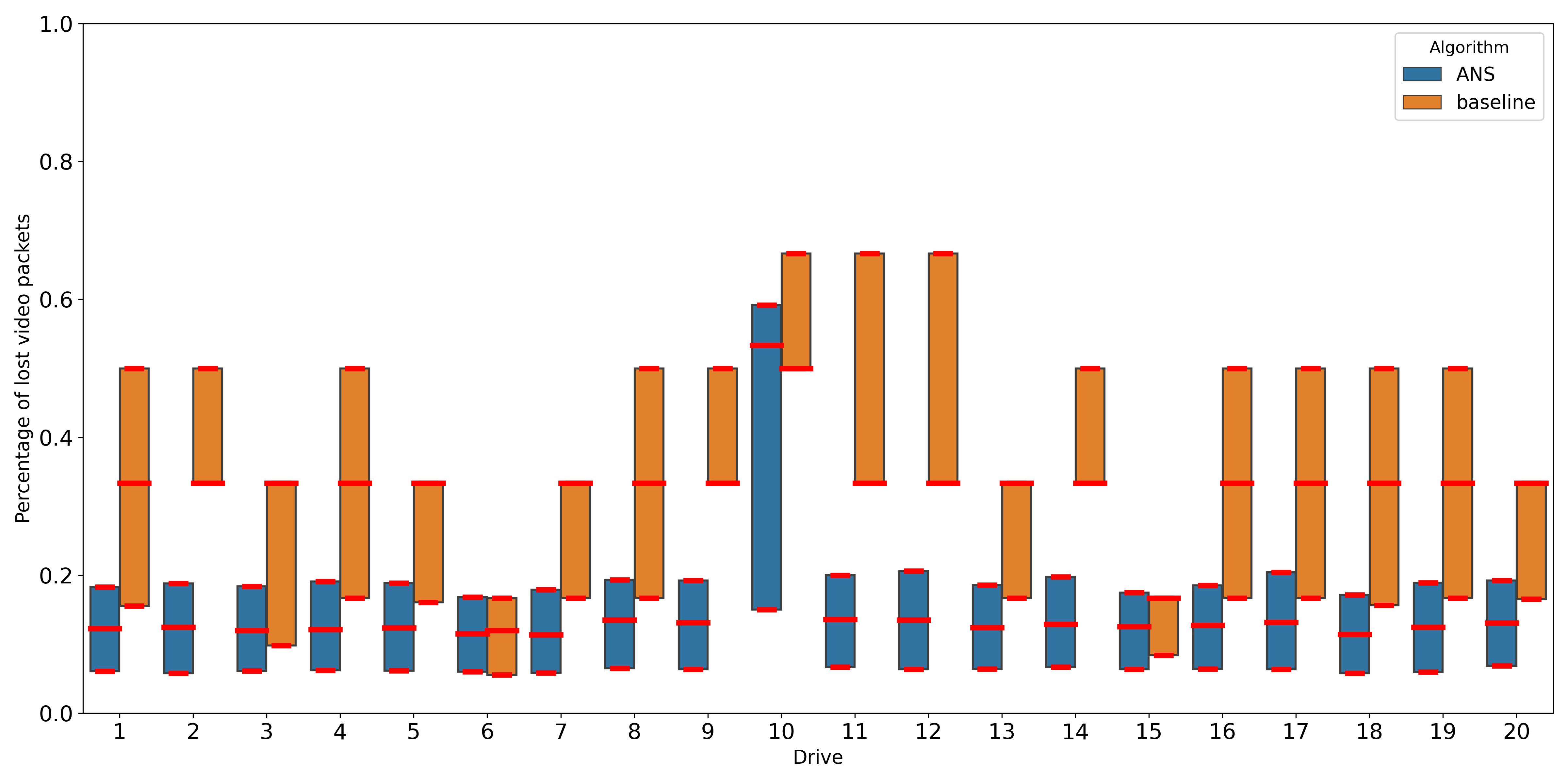}
  \caption{Video packet loss rate of Baseline and ANS across 20 test drives}
  \label{fig:ratio_compare}
\end{figure}%
\begin{algorithm}[!tb]
\centering
\caption{\begin{footnotesize}Active Network Selection (ANS)\end{footnotesize}}
\scalebox{0.7}{ 
\begin{minipage}{1.25\textwidth} 
\begin{algorithmic}[1]
\State Initialize $CN$ as an empty list
\For{each $cellular\_network_i$}
    \State $latency_i \gets$ LatencyPredict($cellular\_network_i$)
    \State $handover\_p_i \gets$ HandPredict($cellular\_network_i$)
    \State Add ($latency_i$,$handover\_p_i$) to $CN$
\EndFor
\State Sort $CN$ from the minimum to the maximum latency
\For{each $cellular\_network$ in sorted $CN$}
    \If{$handover\_prob_i < threshold$}
        \State \text{choose this network for the next packets}
        \State \textbf{exit}
    \EndIf
\EndFor
\State Choose a random network from CN for the next packets
\end{algorithmic}
\end{minipage}
}
\label{ICOS-algo}
\end{algorithm}
We now present our Active Network Selection (ANS) algorithm for solving the MPR problem from Section \ref{INTRO}. This algorithm uses both HandPredict and LatencyPredict to select the cellular network over which the next video packets are transmitted. It first predicts the handover probability for each cellular network, using the unified trained ML model of HandPredict. This step is conducted concurrently across each of the cellular networks. Then, it predicts the latency using LatencyPredict, concurrently for each cellular network. All these predictions are conducted with a one-second time horizon, ten time steps ahead. It then considers only the cellular networks whose handover probability does not exceed a predetermined threshold. From these cellular networks, it chooses the one for which the predicted latency is the minimum; see Algorithm \ref{ICOS-algo} for more details.

ANS prioritizes latency over packet loss because, as we discover, only $3\%$ of the samples have handovers. Therefore, most of the time, the cellular network with the lowest predicted latency is chosen. However, if this cellular network has a high handover probability, we choose the cellular network with the second minimum latency. If all the cellular networks have a high handover probability, the algorithm chooses one of them randomly.

The runtime complexity of ANS includes running two ML models and a constant overhead. We assume a fixed input size of $9 \times 64$, for the 9 features and 64 window slots. For the convolutional layers, the complexity is $O(C_{in} \times C_{out} \times K \times L)$, where $C_{in}$ and $C_{out}$ are the number of input and output channels, respectively. In our case, $C_{in}=9$ for the 9 input features and $C_{out}=64$ for the 64 output neurons selected for the convolution layers. $K$ is the kernel size, which is $3$, and $L$ is the sequence length, which is the number of window slots (64). The complexity of the LSTM layers is $O(N \times T\times (D\times H + H^{2}))$, where $N$ is the batch size (512), $T$ is the number of time steps (64), $H$ is the length of the hidden layers (128), and $D$ is the output channels from the last convolutional layer (128). The complexity of each fully connected layer that follows the LSTM layers is $O(N \times D_{in} \times D_{out})$, where $D_{in}$ and $D_{out}$ represent the input and output dimensions respectively, and $N$ is the batch size. The average running time of LatencyPredict and HandPredict on the 20 test drives was 9ms. Thus, if LatencyPredict and HandPredict are executed sequentially by the algorithm, as in Algorithm \ref{ICOS-algo}, the total running time is 18ms, which is much lower than the 100ms latency constraint.

We compare ANS to the current baseline algorithm used by DriveU, which sends 30 video frames per second across three cellular networks simultaneously in the following way. Each video frame, which originally consists of 24 video packets, is encoded with a FEC scheme. This FEC increases the length of each video frame by 50\%: from 24 to 36 packets. These 36 packets are then evenly distributed across the three networks and any 24 packets are sufficient for reconstructing the original video frame. For the comparison with ANS, the 9-feature model with a 64-second window is used for both HandPredict and LatencyPredict, and both are trained on a dataset of 80 drives.

\begin{figure}[!tb]
\centering
  \captionsetup{justification=centering}
  \includegraphics[width=1\linewidth]{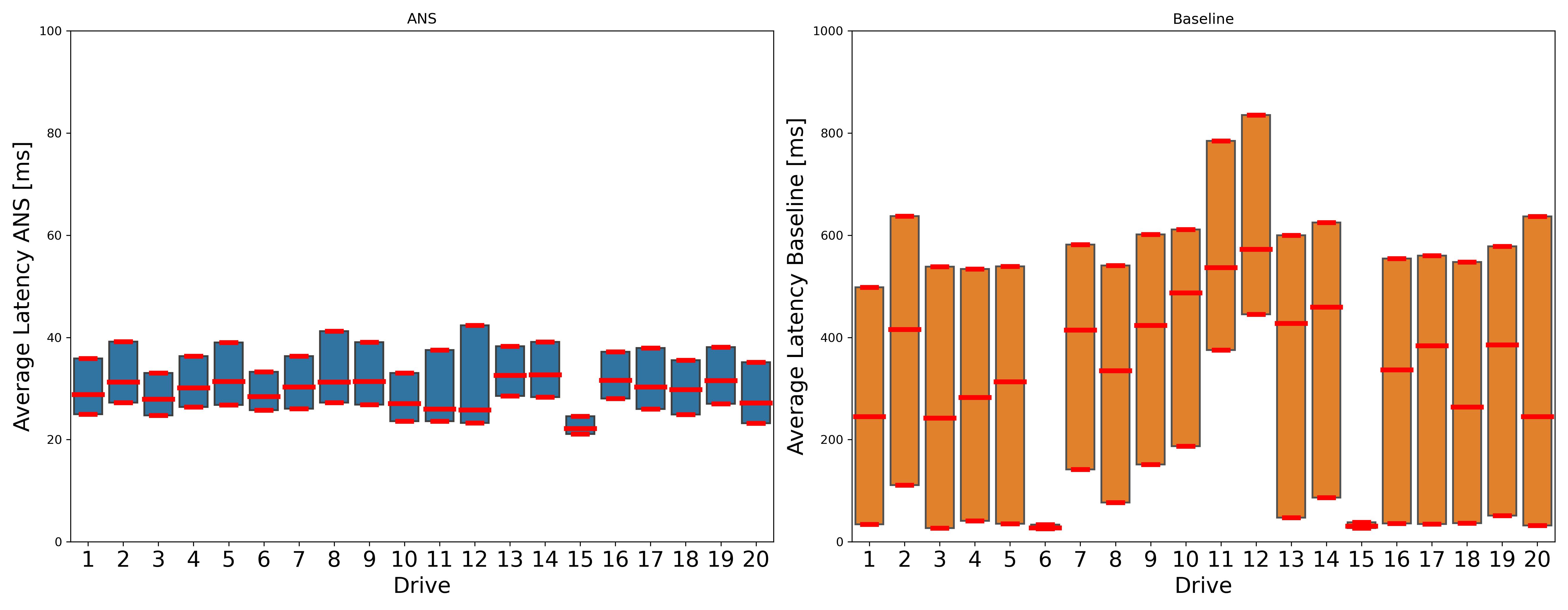}
  \caption{The average latency of each video packet for each test drive}
  \label{fig:latency_comparison}
\end{figure}
Figure \ref{fig:ratio_compare} compares the packet loss rate of Baseline and ANS across 20 different test drives, held in the same location and the same fixed route with challenging cellular communication conditions. We did not use these test drives during training. The x-axis represents these 20 test drives. Each test drive lasts between 900 to 1,200 seconds. Thus, ANS makes between 900 and 1,200 decisions during each drive. The y-axis represents the 25$^\mathrm{th}$, 50$^\mathrm{th}$, and 75$^\mathrm{th}$ percentile distributions of the packet loss rate per second across these test drives. For example, the  blue rectangle for the first test drive indicates that during 25\% of the seconds of the first drive, ANS had a loss rate of at most 8\%; during 75\% of the seconds of this drive, ANS had a loss rate of at most 20\%; and during 50\% of the seconds of this drive, ANS had a loss rate of at most 14\%. Note that in some cases, the median (50$^\mathrm{th}$ percentile) is very close to the 25$^\mathrm{th}$ percentile or to the 75th percentile. Thus, it is represented by a slightly longer line. For example, in the third test drive, in the orange rectangle (for Baseline), the 50$^\mathrm{th}$ is very close to the top of the rectangle (75$^\mathrm{th}$ percentile). It is evident from this graph that ANS performs significantly better than Baseline for all test drives. 

Figure \ref{fig:latency_comparison} compares the average latency per second of the video packets transmitted by the two algorithms. As in Figure \ref{fig:ratio_compare}, this figure also shows the 25$^\mathrm{th}$, 50$^\mathrm{th}$ and 75$^\mathrm{th}$ percentiles. We can see that, with respect to latency, ANS outperforms Baseline on all the test drives in terms of both the variance and the average latency.

%% file: sections/7-conclusion.tex
We defined a new problem, called MPR, for selecting the cellular network over which the next video packets will be transmitted by an AV. To address this problem, a new algorithm, which we called ANS, was presented. ANS aims to minimize the packet loss rate and packet latency, with no prior knowledge of the geographical area, the AV mobility pattern, or the cellular network coverage. It has two ML components: HandPredict for handover prediction and LatencyPredict for packet latency prediction. Using 100 test drives, of which 80 are used for learning and 20 are unseen for testing, we showed that ANS performs much better than a commercial baseline non-learning algorithm.

%% file: root.bbl
\begin{thebibliography}{10}

\bibitem{abdah2020handover}
Hadeel Abdah, Jo{\~a}o~Paulo Barraca, and Rui~L Aguiar.
\newblock {Handover prediction integrated with service migration in 5G systems}.
\newblock In {\em ICC 2020-2020 IEEE International Conference on Communications (ICC)}, pages 1--7. IEEE, 2020.

\bibitem{ahmad2018efficient}
Rami Ahmad, Elankovan~A Sundararajan, Nor~E Othman, and Mahamod Ismail.
\newblock Efficient handover in {LTE}-a by using mobility pattern history and user trajectory prediction.
\newblock {\em Arabian Journal for Science and Engineering}, 43:2995--3009, 2018.

\bibitem{ahn2022comparison}
Hyun Ahn, Kyunghee Sun, and K~Kim.
\newblock {Comparison of missing data imputation methods in time series forecasting}.
\newblock {\em Computers, Materials \& Continua}, 70(1):767--779, 2022.

\bibitem{https://doi.org/10.48550/arxiv.2103.01600}
Parikshit Bansal, Prathamesh Deshpande, and Sunita Sarawagi.
\newblock {Missing Value Imputation on Multidimensional Time Series}, 2021.

\bibitem{chen2013predicting}
Xu~Chen, Fran{\c{c}}ois M{\'e}riaux, and Stefan Valentin.
\newblock Predicting a user's next cell with supervised learning based on channel states.
\newblock In {\em 2013 IEEE 14th workshop on signal processing advances in wireless communications (SPAWC)}, pages 36--40. IEEE, 2013.

\bibitem{chen2018gradnorm}
Zhao Chen, Vijay Badrinarayanan, Chen-Yu Lee, and Andrew Rabinovich.
\newblock Gradnorm: Gradient normalization for adaptive loss balancing in deep multitask networks.
\newblock In {\em International conference on machine learning}, pages 794--803. PMLR, 2018.

\bibitem{chiu2000predictive}
Ming-Hsing Chiu and Mostafa~A Bassiouni.
\newblock {Predictive schemes for handoff prioritization in cellular networks based on mobile positioning}.
\newblock {\em IEEE Journal on selected areas in communications}, 18(3):510--522, 2000.

\bibitem{fattore2020automec}
Umberto Fattore, Marco Liebsch, Bouziane Brik, and Adlen Ksentini.
\newblock {AutoMEC: LSTM-based user mobility prediction for service management in distributed MEC resources}.
\newblock In {\em Proceedings of the 23rd International ACM Conference on Modeling, Analysis and Simulation of Wireless and Mobile Systems}, pages 155--159, 2020.

\bibitem{frank2001time}
Ray~J Frank, Neil Davey, and Stephen~P Hunt.
\newblock Time series prediction and neural networks.
\newblock {\em Journal of intelligent and robotic systems}, 31:91--103, 2001.

\bibitem{GOLAGHAZADEH2022116597}
Firouzeh Golaghazadeh, Stéphane Coulombe, and Jean-Marc Robert.
\newblock Residual packet loss rate analysis of 2-d parity forward error correction.
\newblock {\em Signal Processing: Image Communication}, 102:116597, 2022.

\bibitem{8141873}
Fazle Karim, Somshubra Majumdar, Houshang Darabi, and Shun Chen.
\newblock {LSTM Fully Convolutional Networks for Time Series Classification}.
\newblock {\em IEEE Access}, 6:1662--1669, 2018.

\bibitem{https://doi.org/10.48550/arxiv.2111.13879}
Muhammad~Asif Khan, Ridha Hamila, Adel Gastli, Serkan Kiranyaz, and Nasser~Ahmed Al-Emadi.
\newblock {ML-based Handover Prediction and AP Selection in Cognitive Wi-Fi Networks}, 2021.

\bibitem{kiranyaz20211d}
Serkan Kiranyaz, Onur Avci, Osama Abdeljaber, Turker Ince, Moncef Gabbouj, and Daniel~J Inman.
\newblock 1d convolutional neural networks and applications: A survey.
\newblock {\em Mechanical systems and signal processing}, 151:107398, 2021.

\bibitem{OpenCellID}
OpenCellID.
\newblock Opencellid mapping platform, 2023.
\newblock Accessed: August 11, 2023.

\bibitem{ordonez2016deep}
Francisco~Javier Ord{\'o}{\~n}ez and Daniel Roggen.
\newblock {Deep convolutional and LSTM recurrent neural networks for multimodal wearable activity recognition}.
\newblock {\em Sensors}, 16(1):115, 2016.

\bibitem{patro2015normalization}
SGOPAL Patro and Kishore~Kumar Sahu.
\newblock Normalization: A preprocessing stage.
\newblock {\em arXiv preprint arXiv:1503.06462}, 2015.

\bibitem{peterson2009k}
Leif~E Peterson.
\newblock {K-nearest neighbor}.
\newblock {\em Scholarpedia}, 4(2):1883, 2009.

\bibitem{rehfeld2011comparison}
Kira Rehfeld, Norbert Marwan, Jobst Heitzig, and J{\"u}rgen Kurths.
\newblock Comparison of correlation analysis techniques for irregularly sampled time series.
\newblock {\em Nonlinear Processes in Geophysics}, 18(3):389--404, 2011.

\bibitem{riley2010ns}
George~F Riley and Thomas~R Henderson.
\newblock The ns-3 network simulator.
\newblock {\em Modeling and tools for network simulation}, pages 15--34, 2010.

\bibitem{tan2022intelligent}
Kang Tan, Duncan Bremner, Julien Le~Kernec, Yusuf Sambo, Lei Zhang, and Muhammad~Ali Imran.
\newblock Intelligent handover algorithm for vehicle-to-network communications with double-deep q-learning.
\newblock {\em IEEE Transactions on Vehicular Technology}, 71(7):7848--7862, 2022.

\bibitem{Teleoperation}
Techopedia.
\newblock Autonomous vehicle teleoperation: Is one network enough for remote driving?
\newblock \url{https://www.techopedia.com/definition/15032/teleoperation}, 2017.

\bibitem{tsai2010sub}
Ming-Fong Tsai, Ce-Kuen Shieh, Chih-Heng Ke, and Der-Jiunn Deng.
\newblock Sub-packet forward error correction mechanism for video streaming over wireless networks.
\newblock {\em Multimedia Tools and Applications}, 47:49--69, 2010.

\bibitem{yen2006under}
Show-Jane Yen and Yue-Shi Lee.
\newblock Under-sampling approaches for improving prediction of the minority class in an imbalanced dataset.
\newblock In {\em Intelligent Control and Automation: International Conference on Intelligent Computing, ICIC 2006 Kunming, China, August 16--19, 2006}, pages 731--740. Springer, 2006.

\bibitem{yu2019review}
Yong Yu, Xiaosheng Si, Changhua Hu, and Jianxun Zhang.
\newblock A review of recurrent neural networks: Lstm cells and network architectures.
\newblock {\em Neural computation}, 31(7):1235--1270, 2019.

\end{thebibliography}
